\documentclass[twocolumn,prb,aps,amsmath,amssymb,superscriptaddress,showpacs]{revtex4-2}
\usepackage{dcolumn}
\usepackage{hyperref}
\hypersetup{
    colorlinks=true,
    linkcolor=blue,
    filecolor=magenta,      
    urlcolor=blue,
    citecolor=blue,
}
\usepackage{epsfig}
\usepackage{graphics}
\usepackage{float}
\usepackage{amsmath}
\usepackage{xcolor,colortbl}
\usepackage[utf8]{inputenc}
\usepackage[T1]{fontenc}
\usepackage{lmodern} 
\usepackage{soul}

\definecolor{Gray}{rgb}{0.72,0.72,0.98}
\definecolor{LightCyan1}{rgb}{0.83,0.83,0.98}
\definecolor{LightCyan2}{rgb}{0.91,0.92,1}

\begin{document}
\newcommand{\upn}{Departamento de Ciencias, Universidad Privada del Norte, Lima 15434, Peru}

\newcommand{\cnr} {Istituto di Struttura della Materia of the National Research Council, Via Salaria Km 29.3,I-00016 Monterotondo Stazione, Italy}
\newcommand{\etsf} {European Theoretical Spectroscopy Facilities (ETSF)}
\newcommand{\ift} {Instituto de F\'{\i}sica Te\'{o}rica, Universidade Estadual Paulista (UNESP), Rua Dr. Bento T. Ferraz, 271, S\~{a}o Paulo, SP 01140-070, Brazil.}
\title{Screened hydrogen model of excitons in semiconducting nanoribbons}

\author{Cesar E. P. Villegas}
\affiliation{\upn}
\author{Alexandre R. Rocha}
\affiliation{\ift}


\date{\today}

\begin{abstract}
 The optical response of quasi-one-dimensional systems is often dominated by tightly bound excitons, that significantly influence their basic electronic properties. Despite their importance for device performance, accurately predicting their excitonic effects typically requires computationally demanding many-body approaches. Here, we present a simplified model to describe the static macroscopic dielectric function, which depends only on the width of the quasi-one-dimensional system and its polarizability per unit length. We show that at certain interaction distances, the screened Coulomb potential is greater than its bare counterpart, which results from the enhanced repulsive  electron-electron interactions. 
  As a test case, we study fourteen different nanoribbons, twelve of them armchair graphene nanoribbons of different families. Initially, we devised a simplified equation to estimate the exciton binding energy and extension that provides results comparable to those from the full Bethe-Salpeter equation, albeit for a specific nanoribbon family. Then, we used our proposed screening potential to solve the 1D Wannier-Mott equation, which turn out to be broad approach, that is able to  predict binding energies that match quite well the ones obtained with the Bethe-Salpeter equation, irrespective of the nanoribbon family.
\end{abstract}
\maketitle
\section{Introduction}
Atom-thick quasi-one-dimensional (Q1D) semiconducting systems hold promise as platforms for electronic and optoelectronic applications owing to their tunable band gaps and optical responses that cover a broad range of the electromagnetic spectrum. According to previous theoretical works, it is now well established that Coulomb interactions play a key role in the transport and optical properties of Q1D semiconductors \cite{yang2007,prezzi2008,villegas2014}. More recently, these predictions have also been observed experimentally \cite{tries2020,schmidt2019dielectric}.

Coulomb screening plays a fundamental role in determining a wide range of physical properties in solids and molecular systems \cite{chazalviel1999,wang2020}. In particular, screened Coulomb interactions are critical for the formation of excitonic states. In conventional three-dimensional bulk semiconductors, excitons are weakly bound (a few to tens of meV) due to the large environmental screening \cite{dvorak2013,cardona2005}. However, systems of reduced dimensionality (2D and 1D) exhibit strongly bound excitons as a consequence of the reduced dielectric screening they experience, resulting from significant changes in the dielectric environment \cite{chernikov2014,tries2020}.
Theoretically, the exciton binding energy can be accurately determined by solving the Bethe-Salpeter equation (BSE) within a framework of first-principles methods based on many-body perturbation theory \cite{rohlfing2000,blase2018}. However, this is very computationally demanding and can only be calculated for relatively small systems. Since the enhanced electron-hole interactions in low-dimensional semiconductors considerably impact their optical properties, simplified models for the screened Coulomb potential that are able to predict exciton binding energies and radii for realistic materials are highly desirable. Additionally, these models may provide useful insights that might be hidden by more complex calculations.

In this regard, the simple soft-Coulomb and modified Kratzer potentials have been used as an alternative to include effects of screening in some 2D semiconductors \cite{molas2019energy}, yielding reasonable binding energies for excitons \cite{huang2013,villegas2016}, and interlayer excitons \cite{martins2020}. Moreover, the Rytova-Keldysh potential \cite{rytova1967screened,keldysh1979}, and its extended form for truly 2D semiconductors \cite{cudazzo2011}, which depends on the polarizability of the material, has also been successfully used to account for the electron-hole interaction. 

While considerable theoretical advances have been made in modeling screened potentials in 2D semiconductors, describing electron-hole interactions in quasi-one-dimensional (Q1D) semiconducting systems has typically relied on cusp-type \cite{ogawa1991optical}, Yukawa-like \cite{wang2014}, and soft-Coulomb potentials \cite{bryant1982,brazovskii2010physical,grasselli2017}. Accurate models for electron-hole interaction in carbon nanotubes (CNTs) have been proposed, utilizing either fitted parameters from first principles calculations or two-band models for the dielectric function \cite{deslippe2009,sesti2022}. However, the counterpart for atom-thick 1D semiconducting ribbons is still lacking. 
This is particularly critical given modern synthesis techniques that have enabled the realization of novel atom-thick quasi-one-dimensional (Q1D) nanostructures \cite{cai2010}, where excitonic effects dominate.

In this work, we propose a simple model to describe the static dielectric response and screened Coulomb  potential energy of any atom-thick Q1D semiconducting system. These physical quantities are used to solve the 1D-Wannier-Mott equation and study the excitonic effects of fourteen semiconducting nanoribbons, including 12 armchair graphene nanoribbons of different families, whose exciton extension and binding energies, for the lowest-excitonic state, compares well with calculations based on the solution of the Bethe-Salpeter equation.

\section{Modeling and Discussion}
The screening is modeled starting from the proposal by Cudazzo \emph{et al.} for 2D dielectrics \cite{cudazzo2011}. Our model initially considers an infinitely long narrow dielectric rod that extends along the $x$-direction with width $L$ (along the $y$-direction) and thickness $b$ (along the $z$-direction), fully surrounded by vacuum. Then, a point charge with charge density $\rho^{ext}(\textbf{r})=e\delta(\textbf{r})$ is placed at the origin of the dielectric, causing a redistribution of charges in its surroundings, which is characterized by the induced charge density $\rho^{ind}(\textbf{r})=-\nabla \cdot \textbf{P$_{1D}$}(\textbf{r$_{x}$})$. The induced charge is restricted to the dielectric rod, and it is evaluated at a point $\textbf{r$_{x}$} = (x,|y|\le \frac{L}{2},|z|\le \frac{b}{2})$. The polarization is assumed to be proportional to the induced electric field $\textbf{P$_{1D}$}= \alpha_{1D} \textbf{E}_{local}$, which enables us to rewrite the induced density in terms of the total electrostatic potential, $\rho^{ind}(\textbf{r}_{x})= - \alpha_{1D} \delta(\frac{L}{2}-|y|)\delta(\frac{b}{2}-|z|)\nabla^{2} \phi(\textbf{r}_{x})$, where $\alpha_{1D}$ represents the 1D internal polarizability per unit length of the system. Here, the delta functions centered at the edges of the rod are mathematical artifacts chosen to avoid indeterminacy in certain definite integrals that arise during the Fourier transform procedure. Hence, in the limit of small $L$ and $b$ going to zero, the induced charge density is effectively localized on the surface of the Q1D system. The screened potential is thus obtained by solving the Poisson equation, $\nabla^{2} \phi(\textbf{r}) = -4\pi [ \rho^{ext}(\textbf{r})+\rho^{ind}(\textbf{r})]$, which, after some algebra, allows us to write down the screened electrostatic potential in reciprocal space,
\begin{equation}
\label{eq3}
 \tilde{\phi}_{1D}(q_{x})=\frac{2eK_{0}(\frac{L}{2}|q_{x}|)}{1+8\alpha_{1D}q_{x}^{2}K_{0}(\frac{L}{2}|q_{x}|)}.
\end{equation}
Here $K_0$ represent the zeroth-order modified Bessel function of the second kind. From the latter expression, one can immediately define the 1D microscopic dielectric response function 
\begin{equation}
\label{eq4}
 \varepsilon_{1D}(q_{x})= 1+8\alpha_{1D}q_{x}^{2}K_{0}\left(\frac{L|q_{x}|}{2}\right),
\end{equation}
which is clearly $q_{x}$-dependent, reflecting its nonlocal nature in real space. This expression bears similarity to the dielectric function of single-wall carbon nanotubes obtained within the random phase approximation (RPA), with the exception of a factor that inherently describes the cylindrical geometry of CNTs  \cite{bulashevich2003}. The reader is referred to Appendix A, which contains the full derivation of equations (\ref{eq3}) and (\ref{eq4}).
\begin{figure}[ht]
\centering
\includegraphics[width=1.0\columnwidth]{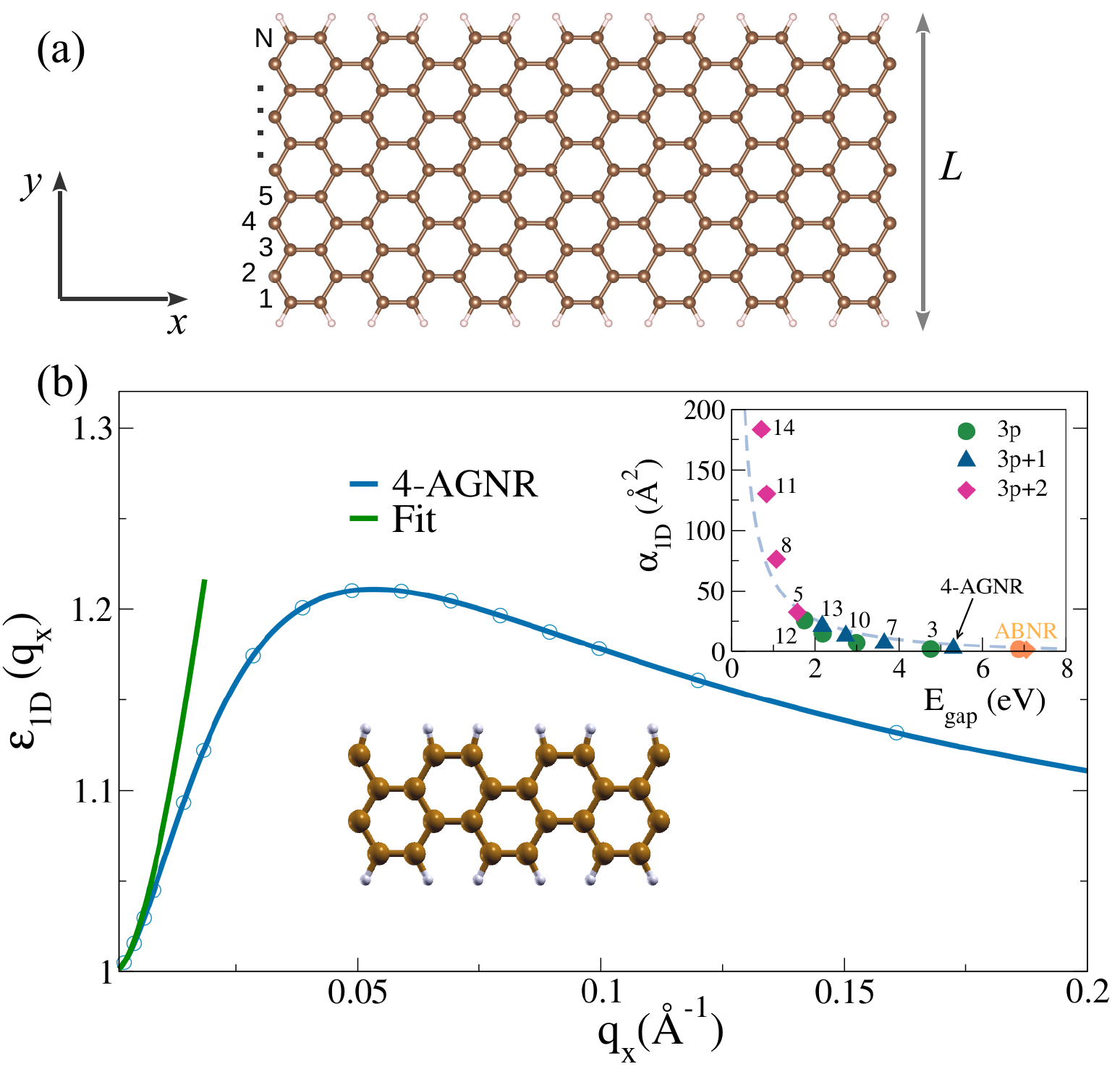}
\caption{\small{(a) Schematic representation of an armchair ribbon indicating the number of dimmer lines across the ribbon width. The ribbon is periodic along the $x$-direction. (b) RPA \emph{ab initio} static dielectric response function for a 4-AGNR (solid blue line). The green line represents the fit to the function $\varepsilon_{1D}(q_{x})\approx 1-8\alpha_{1D}q_{x}^{2}\ln(q_{x}L/4)$. The ribbon width and polarizability per unit length are $L=5.54$ $\AA$ and $\alpha_{1D}=2.59$ $\AA^2$, respectively. The inset shows the dependence of $\alpha_{1D}$ with the quasiparticle band gap of fourteen different armchair ribbons. The dot, triangle, and diamond symbols differentiate the different nanoribbon families, while the indices next to the symbols indicate the nanoribbon index. The orange symbols correspond to selected armchair boron nitride nanoribbons (ABNR) \cite{wang2011optical}.
The gray dashed line represents a fitted reciprocal function used as guide to the eyes.}}\label{fig1} 
\end{figure}

In order to assess the validity of our dielectric screening model, we compare it with many-body quasiparticle calculations obtained from first principles, for a variety of semiconducting armchair graphene nanoribbons (AGNRs) with different lateral confinement sizes (widths). Depending on the number of atoms $N$ that define the lateral width (see Fig. \ref{fig1}a), AGNRs can be classified into three different families named $N$=3$p$, 3$p$+1, and 3$p$+2, where $p$ is an integer. Now it is well established, both theoretically and experimentally, that the three AGNR families are semiconductors and that the band gap is reduced as the lateral width increases, following an inverse relationship \cite{wang2016energy}. Hereafter, we used the nomenclature $N$-AGNR to refer to the systems used in this work.

Figure \ref{fig1}b presents the \emph{ab-initio} calculated $q$-dependent dielectric function of a 4-AGNR obtained within the RPA including local field effects. As expected, in the limit $q$ $\to$ 0, the dielectric function is unity which indicates the absence of long-range screening. For finite momentum transfer, the dielectric function smoothly varies above the unity value, reaching a maximum of approximately 1.2. This indicates that microscopic effects, such as scattering processes are indeed relevant.
Since our semi-classical approach is expected to be valid for small momentum transfer, we can approximate Eq. (\ref{eq4}) as $\varepsilon_{1D}(q_{x})\approx1-8\alpha_{1D}q_{x}^{2}\ln(q_{x}L/4)$ and easily extract $\alpha_{1D}$ by performing a fitting procedure using the \emph{ab initio} calculations. The result of this procedure is also shown in Fig. (\ref{fig1}b) by the green line. In practice, the 1D polarizability of any Q1D system can be roughly estimated by the finite difference relation
\begin{equation}
\label{eq5}
 \alpha_{1D}=\frac{1-\varepsilon(q_1)}{8q_{1}^2\ln(q_{1}L/4)},
\end{equation}
where $q_1$ is a finite but small wave vector that in this work is obtained from a fine uniform $k$-grid sample of 200 $\times$ 1 $\times$ 1. This procedure has already been employed for estimating the polarizability per unit area of 2D systems, yielding reliable results \cite{olsen2016}.
It is also important to analyze the dependence of the 1D polarizability per unit length with the electronic quasiparticle band gap. The inset of Fig. (\ref{fig1}b) shows $\alpha_{1D}$ for 14 nanoribbons of different widths. The 1D polarizability is roughly inversely proportional to the QP band gap except for some deviations occurring for wide band gap systems. This result suggest that Q1D systems may host a universal linear scaling law between the band gap and the exciton binding energy as already observed in 2D systems \cite{jiang2017}.
\begin{figure}[t]
\centering
\includegraphics[width=1.0\columnwidth,height=5.9cm]{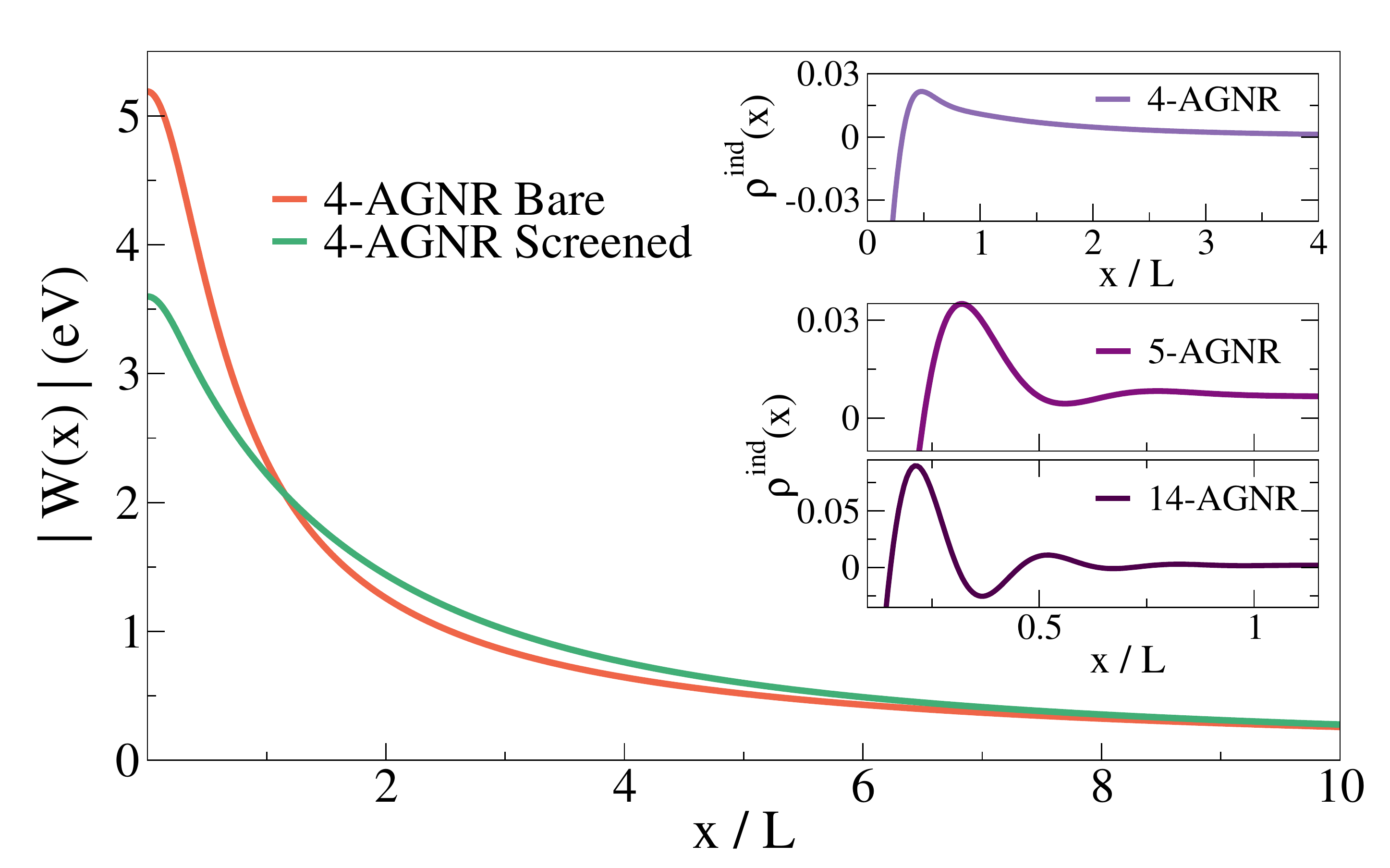}
 \caption{\small {Comparison between the screened and bare energy potential energy for a 4-AGNR. The bare energy potential is obtained by setting $\alpha_{1D}=0$ in Eq. (\ref{eq6}). The insets shows the induced charge density distribution around a point charge at $x=0$. The polarizabilities per unit length for the 5- and 14-AGNR are $\alpha_{1D}$=32.4 $\AA^2$, and $\alpha_{1D}$=183.6 $\AA^2$, respectively.}}\label{fig2} 
\end{figure}

Now, by inverse Fourier transforming Eq. (\ref{eq3}) we obtain the real space screened potential energy, $W_{1D}(x)=e\phi_{1D}(x)$,
\begin{equation}
\label{eq6}
W_{1D}(x)=\frac{e^{2}}{2\pi}\int_{-\infty}^{\infty}dq_{x} e^{iq_{x}x}\frac{2K_{0}(\frac{L}{2}|q_{x}|)}{1+8\alpha_{1D}q_{x}^{2}K_{0}(\frac{L}{2}|q_{x}|)}.
\end{equation}
This is the main result of this work, as it provides a single particle potential for Q1D dielectrics that can be solved numerically with relative care.
In Fig. (\ref{fig2}) we compare the screened Coulomb potential energy of a 4-AGNR with its bare counterpart ($\alpha_{1D}=0$). As expected, at short distances the effective Coulomb potential is considerably reduced by screening effects, which results from attractive interactions between the induced and injected point charge. At large distances, the system is unscreened and follows the characteristic $1/x$ law. This asymptotic trend can be easily demonstrated by analyzing the case of a weakly screened system $\alpha_{1D}\to0$). Indeed, in this limit, the quasi-1D screened potential presents an analytical solution of the form $W_{1D}(r)=e^2[x^2+(L/2)^2]^{-\frac{1}{2}}$, which shows that for long range interactions it follows the characteristic Coulomb potential. Note that this analytical form for the potential energy resembles the widely used potential to describe electron-hole interactions in polymers \cite{brazovskii2010physical}. This particular behavior of our screening model suggest that it can be potentially used to describe polymeric systems more accurately. Interestingly, at intermediate distances, the screened potential slightly rises above the bare one, giving rise to an anti-screening region, which has also been previously seen in molecular systems \cite{van2000} and carbon nanotubes \cite{deslippe2009,sesti2022}. The origin of this phenomenon is related to the enhancement of repulsive interactions in the system. In the supplemental material we show a comparison of the effective electron-electron force for AGNRs of different widths \cite{supp}.
Given its Q1D nature, we argue that, in principle, super Coulombic electron-electron interaction may also play a crucial role in graphene-based electronic waveguides\cite{williams2011gate,zhang2009guided,villegas2010comment,villegas2012}. 

In order to understand this phenomenon, in the upper panel of the inset, we plot the induced charge density around the injected point charge $\rho^{ind}(x)=\frac{e}{2\pi} \int dq_{x}[\varepsilon(q_{x})^{-1}-1]e^{iq_{x}x}$, which exhibits a change of sign and approaches zero at large distances. Hence, when the interaction between the induced charge and the injected charge is sufficiently attractive, this anti-screening effect occurs. 

We also investigate the effect of reducing the electronic band gap (or, equivalently, increasing $\alpha_{1D}$) on the induced charge density. The bottom panels of the inset in Fig. (\ref{fig2}) present $\rho^{ind}(x)$ for 5-AGNR and 14-AGNRs, whose quasi-particle band gaps are 1.57 eV and 0.71 eV, respectively. These values are significantly smaller than the 4-AGNR quasi-particle band gap (5.3 eV). As in the previous case, the total induced charge density integrates to zero; however, the small band gap semiconductors exhibit Friedel-like oscillations, which become more pronounced with decreasing band gap and are expected to dominate in the semimetallic limit \cite{villegas2013}. 
In contrast to metallic systems, where Friedel oscillations arise from the singularities of the integrand that determines $\rho^{ind}(x)$ and its derivatives \cite{lighthill1958}, the physical origin of the charge density oscillation in our semiconducting systems is rather different and can be ascribed to the enhancement of bound charges in the dielectric system (increased polarizability) as the band gap is reduced (see inset of Fig. \ref{fig1}b). This, in turn, originates from a strong repulsion between bound charges and the impurity charge that eventually produces rapidly damped oscillations.

We now proceed to study of the excitonic properties in Q1D systems. Since in practice, one would desire a simple expression that roughly predicts the exciton binding energy with only few electronic structure parameters, we first tackle the problem considering the Wannier-Mott model for a 1D system subject to a bare Coulomb potential that is screened by a constant dielectric function that is dependent on the exciton quantum number \emph{n}. This yields the Rydberg equation (expressed in Hartree units)
\begin{equation}
\label{eq7}
E_{b}^{n}=-\frac{\mu}{2\varepsilon_{n}^2 n^2},
\end{equation}
where $n$ is the exciton quantum number. Following the model proposed by Olsen \emph{et al.} \cite{olsen2016} for 2D semiconductors, we assume the $n$-dependent dielectric to be an average in reciprocal space,
\begin{equation}
\label{eq8}
 \varepsilon_{n}= \frac{2a_{n}}{\pi} \int_{0}^{1/a_{n}}\tilde{\varepsilon}_{1D}(q_{x})dq_{x},
\end{equation}
where $a_{n}$ is the average $n$-dependent effective exciton extension. 
For simplicity, we employ the model dielectric function in the long wavelength limit (q$_{x}$$\to$0), $\tilde{\varepsilon}_{1D}(q_{x})=1-8\alpha_{1D}q_{x}^{2}\ln(q_{x}L/4)$.  This corresponds to the situation from Eq. (\ref{eq4}) in which the electron wavelength is much larger than the characteristic lateral confinement of the Q1D system. Within this approximation, the dielectric function neglects the microscopic effects occurring at short distances. Nevertheless, we should mention that the long wave limit is usually a good starting point to describe many electron-electron interactions in solids and quantum systems, and is widely used as it enables one to deal with rather more simple analytical expressions \cite{dassarma2009,thakur2017dynamical}.
Moreover, the average dielectric function can be related to the 1D exciton extension via the expression 
\begin{equation}
\label{eq9}
a_{n} = \langle |x| \rangle=\frac{3n^{2}\varepsilon_{n}}{2\mu}.
\end{equation}
This relation is obtained by calculating the expectation value of the absolute value of the position operator, taking into account the wave functions of a 1D hydrogen atom (see Appendix for details). This definition enables us to account for the enhanced exciton extension of higher excitonic levels.
After integrating Eq. (\ref{eq8}), and using Eq. (\ref{eq9}) we obtain a simple transcendental equation
\begin{equation}
\label{eq10}
 \varepsilon_{n}^{3}-\frac{2}{\pi}\varepsilon_{n}^{2}-\frac{64\alpha_{1D} \mu^{2}}{81 \pi n^{2}}\left[1+3\ln\left(\frac{6n^{2}\varepsilon_{n}}{\mu L}\right) \right]=0,
\end{equation}
that combined with Eq. (\ref{eq7}), provides the exciton binding energy of any Q1D atom-thick semiconducting system. 
\begin{figure}[ht]
\centering
\includegraphics[width=1.0\columnwidth]{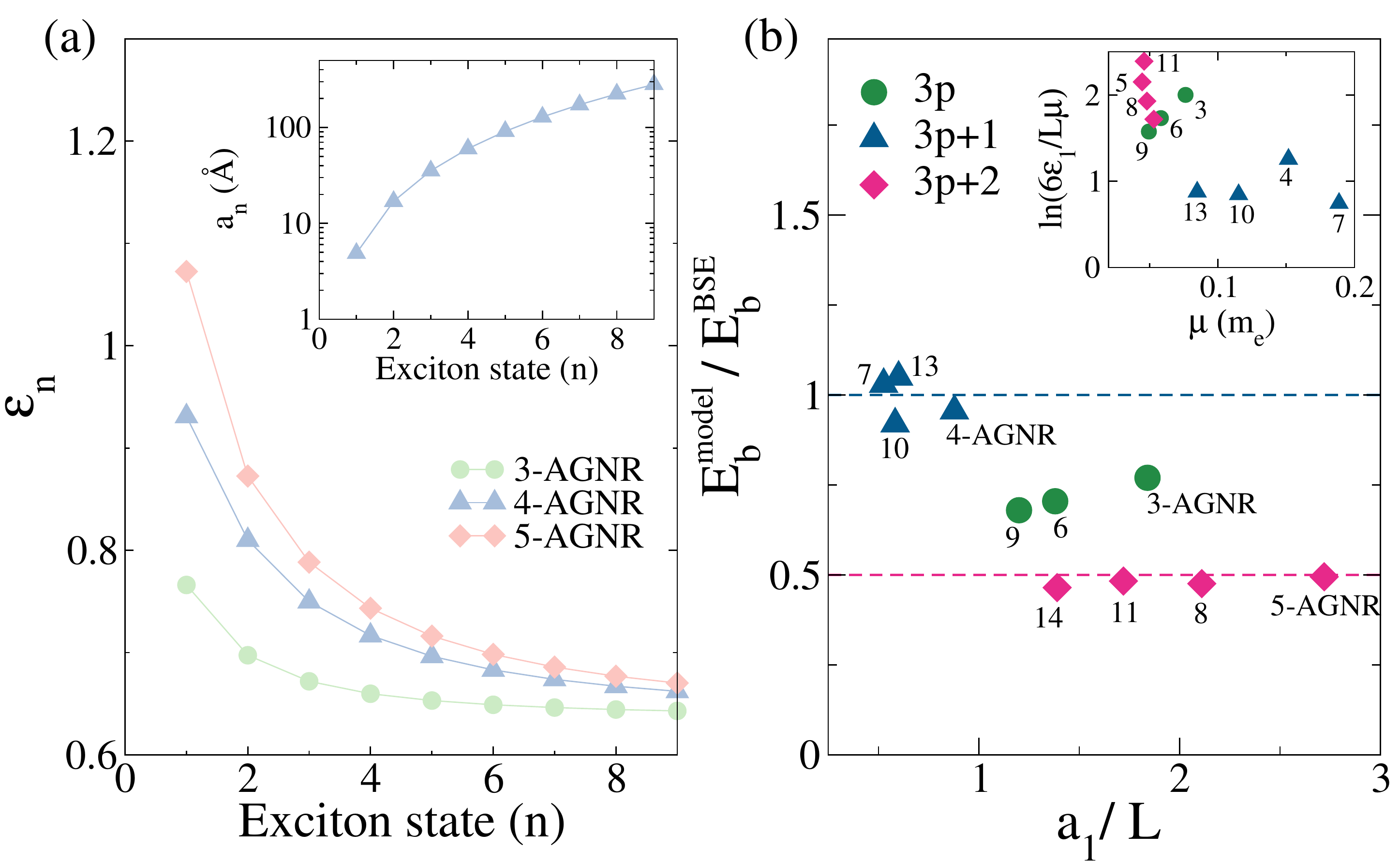}
 \caption{\small {a) $n$-dependent static dielectric function versus the exciton state number $n$ for the 3-, 4-, and 5-AGNRs. The dielectric function is computed by solving Eq. (\ref{eq10}). The inset shows the average exciton extension varying with the exciton state number for the 4-AGNR, which is obtained with Eq. (\ref{eq9}). (b) Binding energy for the lowest-energy bright excitons in twelve different AGNRs as a function of the expectation value of their exciton extension. The labels represent the index of each AGNR. The symbols represent the different AGNR families. The horizontal (vertical) axis is normalized to results obtained by solving the full Bethe-Salpeter equation binding energy (width of the N-AGNR). The inset shows the dependence of the logarithmic factor in Eq. (\ref{eq10}) with the reduced mass.}}\label{fig4} 
\end{figure}

Fig. (\ref{fig4}a) presents the $n$-dependent average dielectric constant for three AGNRs that monotonically decays with the exciton quantum number. For all cases, $\varepsilon_{1}$ yields a value close to unity, which is expected as we are dealing with atom-thick systems surrounded mostly by vacuum. Additionally, one can notice that the factor $n^2\varepsilon^{2}_{n}$ should rapidly enhances with $n$, ensuring the decrease of the binding energy for high level excitonic states (see Eq. (\ref{eq7})). This result translates to enhanced values for the expectation value of the exciton radius for higher $n$, as seen in the inset of Fig.(\ref{fig4}a).
\begin{figure}[th]
\centering
\includegraphics[width=1.0\columnwidth]{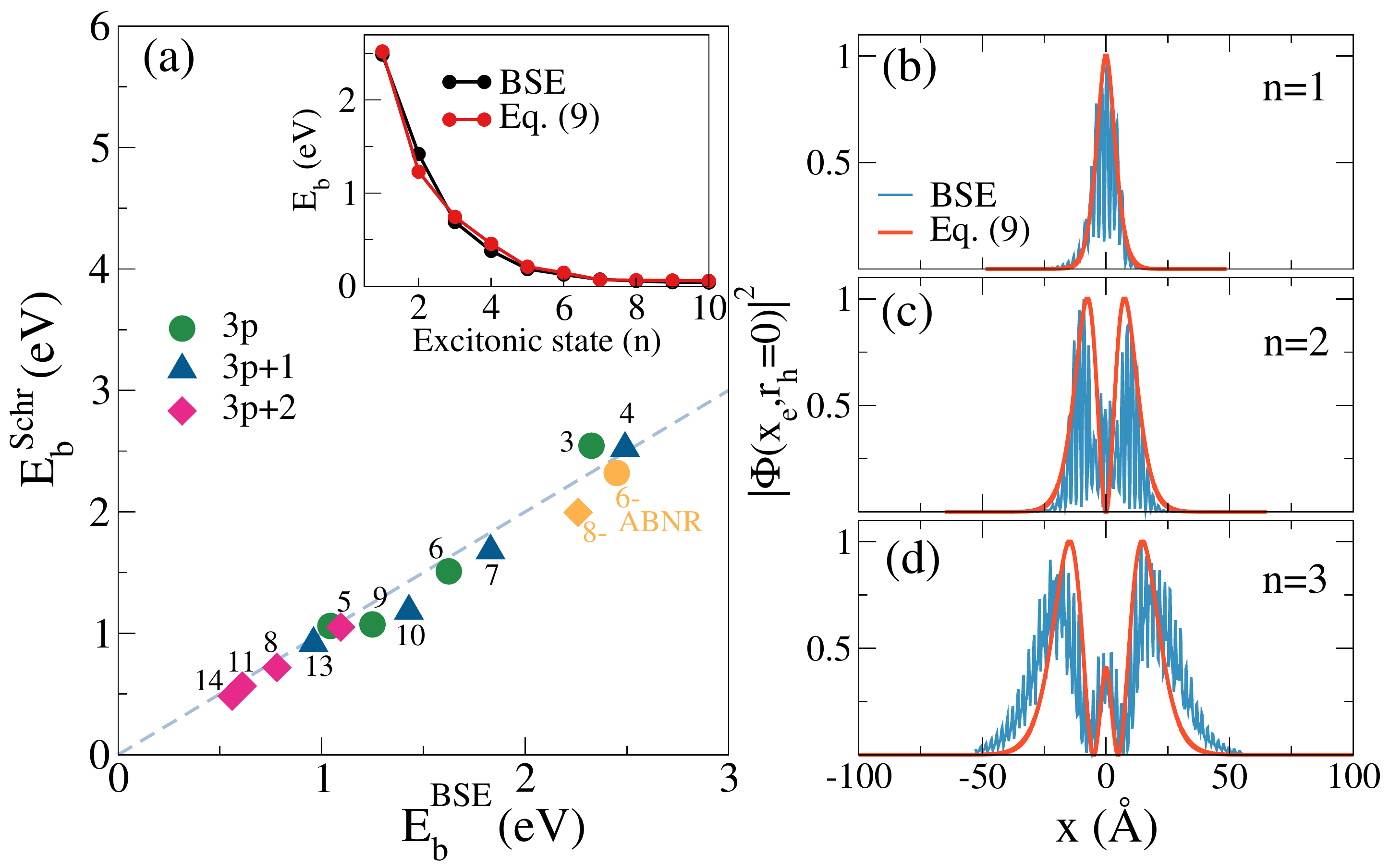}
\caption{\small{(a) Comparison of the lowest-energy 
 bright exciton binding energy calculated with the Bethe-Salpeter equation ($x$-axis) and the 1D-Schr\"{o}dinger equation model given by Eq. (\ref{eq1}) ($y$-axis), for fourteen nanoribbons. The symbols highlight the family of each AGNRs while the labels N=3,4,..12 indicate the width of the ribbon. The inset shows a comparison of the variation of the binding energy with the exciton state number for the BSE (black dots) and Schr\"{o}dinger (red dots) approaches, for the 4-AGNR. The right-handed figure presents a comparison of the normalized electron distribution along the ribbon axis for (b) the first , (c) second, and (d) third bright exciton state of a 4-AGNR.}}\label{fig3} 
\end{figure}

The validity of Eq. (\ref{eq10}) is clearly limited by the behavior of the logarithmic function, which strongly depends on the width and reduced effective mass of the Q1D system. Our simulations show that Eq. (\ref{eq10}) yields real values for $\varepsilon_{n}$ as long as $\ln(6n^{2}/\mu L)\geq 0$. In Fig. (\ref{fig4}b), we present the exciton binding energy for $n=1$, obtained with the simplified model and normalized to results obtained by solving the full Bethe-Salpeter equation as a function of the normalized exciton extension. The binding energies computed within the BSE approach are listed in the supplemental material \cite{supp} (see also references \cite{KohnSham1965,HohenbergKohn1964,qe,van2018,pbe,yambo,bruneval2008,wang2011optical,marinho2024photovoltaic,perez2020} therein). For $a_{1}/L<1$, the simplified screened hydrogen model reproduces the BSE results fairly well. Interestingly, the best agreement occurs for GNRs of the family $3p+1$. This is due to equilibrium between the numerator and denominator of the logarithmic function, which prevent its rapid growth in Eq. (\ref{eq10}). This behavior is shown in the inset of Fig. (\ref{fig4}b) for the three GNR families. For larger values of $a_{1}/L$, however, the model underestimates the exciton binding energies by up to a factor of two for GNRs in the $3p+2$ family. Based on these trends, we expect the analytical screened hydrogen model to reproduce the BSE results for $n=1$ in semiconducting systems where the term $\ln(6n^{2}\varepsilon_{n}/\mu L)$ in Eq. (\ref{eq10}) smoothly varies around unity. It is worth mentioning that the ability of the analytical model to accurately estimate binding energies for different systems may be, in principle, improved by considering explicitly the odd and even parity solutions for the eigenfunctions of the 1D Hydrogen atom Hamiltonian, considering a cut-off potential of the form $V(x)=\frac{e^2}{|x|+\gamma z_{0}}$. Indeed, in this case, the binding energy is proportional to $1/(n+\delta)^2$, where delta depends on the cut-off $\gamma z_{0}$ and the eigenfunctions \cite{haug2009quantum,ogawa1991optical}. However, this procedure precludes one from attaining an analytical expression for the binding energy and will not be addressed in this work.

Although we have shown that using the binding energies obtained from Eq. (\ref{eq10}) we are able to predict reasonably well the BSE results under some constraints, we now turn to a still simple yet more accurate description to predict the exciton binding energy of Q1D semiconductors. 
  
Now we consider the Wannier model for excitons \cite{wannier1937,haug2009quantum}, which relies on the effective mass approximation for an electron-hole pair. The exciton binding energy of a quasi-one dimensional system can be obtained by solving the one-dimensional Schr\"{o}dinger equation,
\begin{equation}
\label{eq1}
\left[\frac{-\hbar^{2}}{2\mu}\frac{\partial^2{}}{\partial^2{x}}+W_{1D}(x)\right]\Psi(x) = E_{b}^{n}\Psi(x),
\end{equation}
where $\mu=m_{e}m_{h}/(m_{e}+m_{h})$ is the reduced effective mass of the electron-hole pair, and $W_{1D}(x)$ represents the real-space screened electron-hole potential defined in Eq. (\ref{eq6}).
 
 In Fig. (\ref{fig3}a), we compare the lowest-energy excitonic state obtained by fully \emph{ab initio} solving the Bethe-Salpeter equation with the results obtained via the numerical solution of Eq. (\ref{eq1}) for 14 different nanoribbons \cite{supp}.  Overall a very good agreement is observed for the entire range of energies for the twelve AGNRS. We also tested our model in two armchair hexagonal boron nitride nanoribbons (ABNR), namely 6-ABNR and 8-ABNR, identified in Fig. (\ref{fig3}a) by the orange symbols and available in literature \cite{wang2011optical}. This result suggests that our model may be potentially used in different planar Q1D semiconducting systems. To explore the predictability of our model for excited states, in the inset of Fig. (\ref{fig3}a) we compare the 10 lowest excitonic states obtained with our model with those obtained from BSE calculations for the 4-AGNR. From this comparison we noted that the model can also describe excited states binding energies.
 
In order to quantitatively assess the extending range of excitons, in Fig. (\ref{fig3}b-d), we present the projected electron density along the ribbon obtained from the solution of the BSE and compare it with the envelope wave function obtained from Eq. (\ref{eq1}). For the two lowest excitonic states, the model of Eq. (\ref{eq1}) is in accordance with the BSE results, and for $n=3$, the model is in qualitative agreement but slightly underestimates the exciton extension by approximately 15 \%. These results suggest that the proposed screening model is able to fairly reproduce the BSE results for the lowest excitonic states based solely on the reduced effective mass and the material dependent 1D polarizability per unit length. These two quantities can be accurately calculated at a modest computational cost using standard density functional theory methods.

At this point, we would like to highlight that our results for the real-space screened potential share some striking similarities with those of semiconducting carbon nanotubes. In these systems, for instance, it has been theoretically predicted that the strength of the Coulomb interactions is considerably enhanced for carrier separation distances larger than the nanotube diameter \cite{capaz2006diameter,sesti2022}. This gives rise to the so-called super-Coulombic interactions, a term coined to refer to the force between two carriers at a given distance, which is stronger than the force associated with Coulomb's law for two electrons in vacuum. 
Interestingly, our results for semiconducting AGNRs point in a similar direction, suggesting that super-Coulombic interactions may also be observed in graphene nanoribbons. Although the experimental verification of super-Coulombic interactions is challenging, as it requires ultra-clean samples and the design of potential traps based on multiple gates, experimental measurements have confirmed the existence of enhanced forces between electrons in carbon nanotubes. These measurements have shown that in certain regions of space, the measured force can be up to 5 times greater than the force predicted by Coulomb's law \cite{shapir2019imaging}. Indeed, this experimental verification opens new avenues for understanding the role of elementary charged interactions in Q1D systems, which may have significant implications for future research in this field.

Another important aspect that our model can address with modest computational cost is the characterization of the width and reduced mass dependence of binding energies for nanoribbons. While this characterization has already been done for AGNRs \cite{zhu2011scaling}, we argue that our model may be employed to perform similar characterizations in other atom-thick nanoribbons, such as boron nitride, silicene, and others. For instance, previous studies conducted on CNTs have characterized the chirality and diameter dependence of exciton binding energies, shedding light on the crucial role of screening effects in excitonic properties \cite{capaz2006diameter,jorio2005resonance}. We argue that characterizing these properties can serve as a valuable guide for interpreting future experimental measurements.

 Finally, we stress that in our study we have considered free-standing systems, although real quasi-one-dimensional (Q1D) semiconductors are typically surrounded by polarized environments such as metal or semiconducting substrates. While the effect of substrates has not been addressed in detail in our study, it is expected that surrounding dielectric materials will considerably reduce the binding energies \cite{denk2014,wang2016energy}. In this regard, we mention that our model can be extended to consider substrate effects following the procedure recently proposed by Riis-Jensen \textit{et al.} \cite{riis2020}.
 \section{Conclusions}
 In Summary, we proposed a simple dielectric screening model for Q1D semiconductors that yield an analytical expression for the macroscopic dielectric function in reciprocal space which solely depends on the effective reduced mass, the system width, and the polarizability per unit length. Based on a screened hydrogen model, we obtained a simple equation for the level dependent dielectric function $\varepsilon_{n}$ and expectation value of the exciton radius $a_{n}$ which describes the excitonic effects of nanoribbons of the $3p+1$ family. Moreover, the real space screening potential is used to solve a 1D Schr\"{o}dinger equation to obtain exciton binding energies and wavefunctions, whose values are in good agreement with those obtained from the solution of an \textit{ab initio}-based Bethe-Salpeter equation.  We expect that this work stimulate future studies on the characterization of excitonic effects in emergent Q1D semiconducting systems, and may serve as a toy model for interpreting future experimental measurements where Coulomb interactions are crucial.
 
 \section{Acknowledgements}
The authors would like to thank Dr. Andrea Marini for fruitful discussions and critical comments. ARR acknowledges support from FAPESP grant numbers 2016/01343-7, 2017/02317-2, 2021/14335-0, 2023/1851-9 and CNPQ grant number 408038/2022-5 and 307710/2019-0. This work used the computational resources of CENAPAD-SP and GRID-UNESP.
\begin{figure}[t]
\centering
\includegraphics[width=1.0\columnwidth]{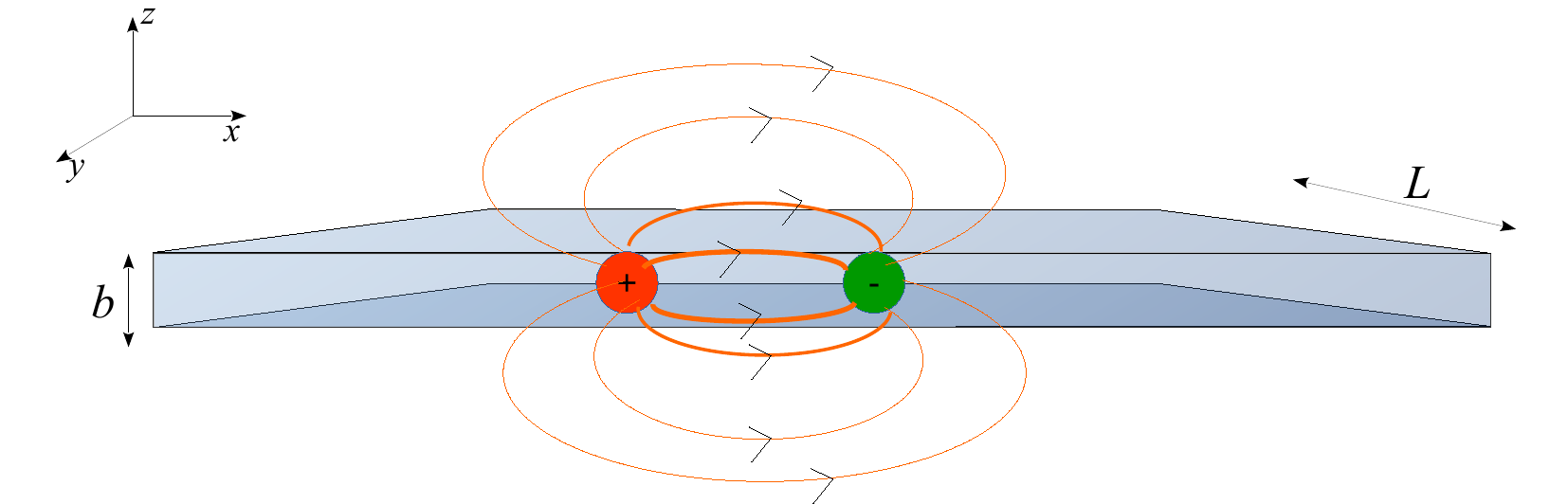}
\caption{\small{Schematic representation of the quasi-one-dimensional system for modelling the screening. Here, we consider an electron (-) and hole (+) in a ribbon, extending along the $x$-direction, of width $L$ and thickness $b$.}}\label{figs0} 
\end{figure}
\section{Appendix}
\subsection{Dielectric screening model for quasi-one-dimensional dielectrics}
Let's consider the effect of inserting an impurity point charge into a Q1D dielectric system represented in Fig. (\ref{figs0}). In response to the electric field associated with the impurity, the local electrical charges slightly reorganize, leading to the polarization of the system. Therefore, one can distinguish between the impurity charge density $\rho^{ext}(\textbf{r})$ and the induced charge density $\rho^{ind}(\textbf{r})$.

The screening is modeled by considering an effective charge density of the form \cite{mahan},
\begin{equation}
\label{eqs1}
 \rho(\textbf{r}) = \rho^{ext}(\textbf{r}) + \rho^{ind}(\textbf{r}).
\end{equation}
Within the linear response regime, the dielectric function is defined as the ratio of the Fourier transformed displacement and electric fields, $\varepsilon(q)=\tilde{D}(q)/\tilde{E}(q)$ along the $\textbf{q}$-direction. From this definition and considering the relation between scalar potentials and the fields, we obtain:
\begin{equation}
\label{eqs4}
 \tilde{\phi}(\textbf{q})=\frac{\tilde{\phi}^{ext}(\textbf{q})}{\epsilon(\textbf{q})},
\end{equation}
Here, $\tilde{\phi}(\textbf{q})$ and $\tilde{\phi}^{ext}(\textbf{q})$ are the Fourier transforms of the screened and impurity scalar potentials, respectively.
Hereafter, we adopt Gaussian units, unless otherwise stated. Now we solve Poisson's equation,
\begin{equation}
\label{eqs5}
 \nabla^{2} \phi(\textbf{r}) = -4\pi \rho(\textbf{r}) =-4\pi \big[\rho^{ext}(\textbf{r}) +\rho^{ind}(\textbf{r})\big], 
\end{equation}
where the impurity charge density is placed at the origin and expressed as,
\begin{equation}
\label{eqs6}
 \rho^{ext}(\textbf{r})=e\delta(\textbf{r}).
\end{equation}

Here, we assume that the induced charge density is $\rho^{ind}(\textbf{r})=-\nabla \cdot \textbf{P}(\textbf{r})$,  while the polarization is proportional to the local electric field $\textbf{P$_{1D}$}= \alpha_{1D} \textbf{E}_{local}$.
The system is initially modeled as a narrow dielectric rod extending in the $x$-direction with width $L$ along the $y$-direction, and thickness $b$ along the $z$-direction. The induced charge density is rewritten in terms of the screened potential as:
\begin{equation}
\label{eqs9}
 \rho^{ind}(\textbf{r$_{x}$})=\alpha_{1D}\delta(\frac{L}{2}-|y|)\delta(\frac{b}{2}-|z|) \nabla^{2}_{x} \phi(\textbf{r$_{x}$}),
\end{equation}
here, $\alpha_{1D}$ represents the dielectric's polarizability per unit length. It's important to note that the induced charge density is confined to the dielectric rod and evaluated at a point $\textbf{r$_{x}$} = (x,|y|\le \frac{L}{2},|z|\le \frac{b}{2})$. The delta functions centered at the edges of the rod serve as mathematical tools to avoid indeterminacy in certain definite integrals encountered during the Fourier transform procedure. It's noteworthy that in the limit of small $L$ and $b$ approaching zero, the induced charge density effectively localizes at the center of the Q1D system.

By introducing Eq. (\ref{eqs6}) and Eq. (\ref{eqs9}) into Poisson's equation, we arrive at the following expression,
\begin{equation}
\begin{split}
 \nabla^{2}\phi=-4\pi e\delta(\textbf{r}) -4\pi\alpha_{1D}\delta(\frac{b}{2}-|z|)\delta(\frac{L}{2}-|y|)\frac{\partial^{2} \phi }{\partial^{2} x},
\end{split}
\end{equation}
which after applying the Fourier transform ($\mathcal{F}$), can be written as
\begin{equation}
\label{eqs11}
\begin{split}
\mathcal{F}[\nabla^{2}\phi)]=-4\pi e \mathcal{F}[\delta(\textbf{r})]-4\pi \alpha_{1D} \mathcal{F}\big[\delta(\frac{b}{2}-|z|)\delta(\frac{L}{2}-|y|)\times \\ \frac{\partial^{2} \phi }{\partial^{2} x}\big].
\end{split}
\end{equation}
The last term on the right-hand side of Eq. (\ref{eqs11}) can be worked out separately 
\begin{widetext}
\begin{equation}
\label{eqs12}
\begin{split}
\mathcal{F}\big[\delta(\frac{b}{2}-|z|)\delta(\frac{L}{2}-|y|)\frac{\partial^{2} \phi }{\partial^{2} x}\big] = -q_{x}^{2} \big\{e^{-q_{z}\frac{b}{2}}e^{-q_{y}\frac{L}{2}} \times  \mathcal{F}_{x}[\phi(x,L/2,b/2)]+e^{-q_{z}\frac{b}{2}}e^{q_{y}\frac{L}{2}}\mathcal{F}_{x}[\phi(x,-L/2,b/2)]+ \ \ \ \ \ \ \ \ \ \ \ \ \ \ \ \ \ \ \ \\ e^{q_{z}\frac{b}{2}}e^{-q_{y}\frac{L}{2}}\mathcal{F}_{x}[\phi(x,L/2,-b/2)]+e^{q_{z}\frac{b}{2}}e^{q_{y}\frac{L}{2}}\mathcal{F}_{x}[\phi(x,-L/2,
-b/2)]\big\}.
\end{split}
\end{equation}
\end{widetext}
This expression can be simplified further by recognizing that a symmetric charge distribution (as assumed in this model) corresponds to a symmetric potential. This allows us to employ the relationships:
$\phi(x,L/2,b/2)=\phi(x,-L/2,-b/2)=\phi(x,-L/2,b/2)=\phi(x,L/2,-b/2)$. Hence, Eq.(\ref{eqs11}) can be expressed as
\begin{equation}
\begin{split}
\label{eq12}
\mathcal{F}[\nabla^{2}\phi)]=-4\pi e \mathcal{F}\delta(\textbf{r})]+16\pi \alpha_{1D} \cos(q_{y}\frac{L}{2})\cos(q_{z}\frac{b}{2}) \times \\ \mathcal{F}\big[ \phi(x,\frac{L}{2},\frac{b}{2})\big]. 
\end{split}
\end{equation}

For very narrow and atom-thick systems, we consider the limit as $L$ and $b$ approach zero. Therefore, we perform a Taylor expansion of the sinusoidal terms and retain only the lowest-order term, which yields:
\begin{equation}
 \begin{split}
 \label{eqs13}
\mathcal{F}[\nabla^{2}\phi)]=-4\pi e \mathcal{F}[\delta(\textbf{r})]+16\pi\alpha_{1D}q^{2}_{x}\mathcal{F}_{x}\big[\phi(x,\frac{L}{2},\frac{b}{2})\big].
 \end{split}
\end{equation}
By now performing the Fourier transform of Eq. (\ref{eqs13}),
\begin{equation}
\label{eqs25}
 (q_{x}^{2}+q_{y}^{2}+q_{z}^{2})\tilde{\phi}(\textbf{q}) \approx 4\pi[ e - 4\alpha_{1D} q_{x}^{2}\tilde{\phi}_{1D}(q_{x})],
\end{equation}
where the one-dimensional Fourier transform of the screened potential, appearing on the left-hand side of Eq. (\ref{eqs25}) adopts the form,
\begin{equation}
\label{eqs26}
\tilde{\phi}_{1D}(q_{x})= \frac{1}{(2\pi)^{2}} \int_{-\infty}^{\infty}dq_{z}dq_{y} e^{i\frac{L}{2}q_{y}} e^{i\frac{b}{2}q_{z}}\tilde{\phi}(\textbf{q}).
\end{equation}
Following a similar procedure as in the 2D case \cite{cudazzo2011}, one can isolate $\tilde{\phi}(\textbf{q})$ from Eq. (\ref{eqs25}) to get,
\begin{equation}
\label{eqs27}
 \tilde{\phi}(\textbf{q})\approx \frac{A(q_{x})+4\pi e}{q_{x}^{2}+q_{y}^{2}+q_{z}^{2}},
\end{equation}
where $A(q_{x})=-16\pi\alpha_{1D}q_{x}^{2}\tilde{\phi}_{1D}(q_{x})$ is defined as a function that depends exclusively on the momentum transferred along the wire. By introducing equation (\ref{eqs27}) into Eq. (\ref{eqs26}) and keeping only the integrals that provide finite values, \textit{i.e}, those containing odd functions, we obtain an expression for the 1D Fourier transform potential
\begin{widetext}
\begin{equation}
\label{eqs27-1}
  \tilde{\phi}_{1D}(q_{x})= \frac{A(q_{x})}{4\pi^{2}}\int^{\infty}_{0}4\cos(\frac{L}{2}q_{y})dq_{y}\int^{\infty}_{0}\frac{\cos(\frac{b}{2}q_{z})}{q_{x}^{2}+q_{y}^{2}+q_{z}^{2}}dq_{z} \\
  +\frac{e}{\pi}\int^{\infty}_{0}4\cos(\frac{L}{2}q_{y})dq_{y}\int^{\infty}_{0}\frac{\cos(\frac{b}{2}q_{z})}{q_{x}^{2}+q_{y}^{2}+q_{z}^{2}}dq_{z}.  
\end{equation}
\end{widetext}
Considering the following identity
\begin{equation}
 \int^{\infty}_{0}\frac{\cos(\frac{b}{2}q_{z})}{q_{x}^{2}+q_{y}^{2}+q_{z}^{2}}dq_{z}=\frac{\pi}{2\sqrt{q_{x}^{2}+q_{y}^{2}}}e^{\frac{b}{2}\sqrt{q_{x}^{2}+q_{y}^{2}}}\approx \frac{\pi}{2\sqrt{q_{x}^{2}+q_{y}^{2}}},
\end{equation}
which is valid in the limit of atom-thick systems ($b$ $\to$ 0). Then equation (\ref{eqs27-1}) adopts the form
\begin{widetext}
\begin{equation}
\begin{split}
  \tilde{\phi}_{1D}(q_{x})= \frac{A(q_{x})}{2\pi}\int^{\infty}_{0}\frac{\cos(\frac{L}{2}q_{y})dq_{y}}{\sqrt{q_{x}^{2}+q_{y}^{2}}}dq_{y}+2e\int^{\infty}_{0}\frac{\cos(\frac{L}{2}q_{y})dq_{y}}{\sqrt{q_{x}^{2}+q_{y}^{2}}} \\
  \tilde{\phi}_{1D}(q_{x}) \approx \frac{A(q_{x})}{2\pi}K_{0}(\frac{L}{2}|q_{x}|)+2eK_{0}(\frac{L}{2}|q_{x}|) = \frac{-16\pi \alpha_{1D}q_{x}^{2}\tilde{\phi}_{1D}(q_{x})}{2\pi}K_{0}(\frac{L}{2}|q_{x}|)+2eK_{0}(\frac{L}{2}|q_{x}|). 
  \end{split}
\end{equation}
\end{widetext}
Now solving for $\tilde{\phi}_{1D}(q_{x})$, we arrive to the expression for the one-dimensional screened potential
\begin{equation}
\label{eqs28}
 \tilde{\phi}_{1D}(q_{x})=\frac{2eK_{0}(\frac{L}{2}|q_{x}|)}{1+8\alpha_{1D}q_{x}^{2}K_{0}(\frac{L}{2}|q_{x}|)}.
\end{equation}
Here $L$ is the effective length scale that characterizes the lateral confinement of a Q1D system and $K_{0}$ is the zeroth-order modified Bessel function of the second kind.
From Eq. (\ref{eqs28}), we notice that the term in the numerator is the quasi-1D Fourier transform of the impurity point charge potential \cite{li1989,dassarma2009}. Thus, a direct comparison of Eq. (\ref{eqs28}) with Eq. (\ref{eqs4}) allows us to identify the dielectric function of a Q1D system,
\begin{equation}
\label{eqs29}
 \epsilon_{1D}(q_{x})= 1+8\alpha_{1D}q_{x}^{2}K_{0}(\frac{L}{2}|q_{x}|),
\end{equation}
as presented in the main text.
\subsection{Expectation value of the position operator}
Consider the Schr\"{o}dinger equation for an electron in a one-dimensional singular Coulomb potential expressed in atomic units \cite{loudon2015}
\begin{equation}
\label{eqs30}
\frac{-1}{2}\frac{d^2{}}{dx^{2}}\Psi(x)-\frac{1}{|x|}\Psi(x) = E\Psi(x),
\end{equation}
whose general solution can be written as \cite{rakhmanov2015},
\begin{equation}
\label{eqs31}
 \Psi_{n}(x)=\sqrt{\frac{2}{n^{3}}}|x|(sgn x)\exp(\frac{-|x|}{n}){}_{1}F_{1}(1-n,2,\frac{2|x|}{n})
\end{equation}
where ${}_1F_{1}$ is the confluent hyper-geometric function.

The expectation value of the absolute value of the position operator $\langle |x| \rangle$ is the key quantity to characterize the extension of the wavefunction. This can be computed as,
\begin{equation}
\label{eqs32}
 \langle |x| \rangle = \langle \Psi_{n}(x)||x||\Psi_{n}(x)\rangle = \int_{-\infty}^{\infty} \Psi_{n}(x) |x| \Psi_{n}^{*}(x)dx.
\end{equation}
Note that we do not compute $\langle x \rangle$ as it is formally zero for any odd or even function. Considering Eq. (\ref{eqs31}) into Eq. (\ref{eqs32}), we obtain the following expression
\begin{equation}
 \langle |x| \rangle = \frac{3n^{2}}{2},
\end{equation}
which is expressed in Bohr radius units.If one is interested in considering the interaction between an electron and hole with reduced mass $\mu$ in a dielectric environment of permittivity $\epsilon$ then,  Eq. (\ref{eqs30}) should be solved by performing the following changes, $m_{e}\to \mu$, and $\frac{e^{2}}{|x|}\to \frac{e^{2}}{ \epsilon|x|}$.  Under this considerations, the expression for the norm of the expectation value of the position yields, 
\begin{equation}
 \langle |x| \rangle = \frac{3n^{2}\epsilon}{2\mu}.
\end{equation}
This expression will be used to quantify the expectation value of the exciton extension in Q1D systems.

\bibliographystyle{apsrev}

\begin{thebibliography}{63}
\expandafter\ifx\csname natexlab\endcsname\relax\def\natexlab#1{#1}\fi
\expandafter\ifx\csname bibnamefont\endcsname\relax
  \def\bibnamefont#1{#1}\fi
\expandafter\ifx\csname bibfnamefont\endcsname\relax
  \def\bibfnamefont#1{#1}\fi
\expandafter\ifx\csname citenamefont\endcsname\relax
  \def\citenamefont#1{#1}\fi
\expandafter\ifx\csname url\endcsname\relax
  \def\url#1{\texttt{#1}}\fi
\expandafter\ifx\csname urlprefix\endcsname\relax\def\urlprefix{URL }\fi
\providecommand{\bibinfo}[2]{#2}
\providecommand{\eprint}[2][]{\url{#2}}

\bibitem[{\citenamefont{Yang et~al.}(2007)\citenamefont{Yang, Cohen, and
  Louie}}]{yang2007}
\bibinfo{author}{\bibfnamefont{L.}~\bibnamefont{Yang}},
  \bibinfo{author}{\bibfnamefont{M.~L.} \bibnamefont{Cohen}}, \bibnamefont{and}
  \bibinfo{author}{\bibfnamefont{S.~G.} \bibnamefont{Louie}},
  \bibinfo{journal}{Nano Lett.} \textbf{\bibinfo{volume}{7}},
  \bibinfo{pages}{3112} (\bibinfo{year}{2007}).

\bibitem[{\citenamefont{Prezzi et~al.}(2008)\citenamefont{Prezzi, Varsano,
  Ruini, Marini, and Molinari}}]{prezzi2008}
\bibinfo{author}{\bibfnamefont{D.}~\bibnamefont{Prezzi}},
  \bibinfo{author}{\bibfnamefont{D.}~\bibnamefont{Varsano}},
  \bibinfo{author}{\bibfnamefont{A.}~\bibnamefont{Ruini}},
  \bibinfo{author}{\bibfnamefont{A.}~\bibnamefont{Marini}}, \bibnamefont{and}
  \bibinfo{author}{\bibfnamefont{E.}~\bibnamefont{Molinari}},
  \bibinfo{journal}{Phys. Rev. B} \textbf{\bibinfo{volume}{77}},
  \bibinfo{pages}{041404} (\bibinfo{year}{2008}).

\bibitem[{\citenamefont{Villegas et~al.}(2014)\citenamefont{Villegas,
  Mendon{\c{c}}a, and Rocha}}]{villegas2014}
\bibinfo{author}{\bibfnamefont{C.~E.~P.} \bibnamefont{Villegas}},
  \bibinfo{author}{\bibfnamefont{P.~B.} \bibnamefont{Mendon{\c{c}}a}},
  \bibnamefont{and} \bibinfo{author}{\bibfnamefont{A.~R.} \bibnamefont{Rocha}},
  \bibinfo{journal}{Sci. Rep.} \textbf{\bibinfo{volume}{4}}, \bibinfo{pages}{1}
  (\bibinfo{year}{2014}).

\bibitem[{\citenamefont{Tries et~al.}(2020)\citenamefont{Tries, Osella, Zhang,
  Xu, Ramanan, Kl\"{a}ui, Mai, Beljonne, and Wang}}]{tries2020}
\bibinfo{author}{\bibfnamefont{A.}~\bibnamefont{Tries}},
  \bibinfo{author}{\bibfnamefont{S.}~\bibnamefont{Osella}},
  \bibinfo{author}{\bibfnamefont{P.}~\bibnamefont{Zhang}},
  \bibinfo{author}{\bibfnamefont{F.}~\bibnamefont{Xu}},
  \bibinfo{author}{\bibfnamefont{C.}~\bibnamefont{Ramanan}},
  \bibinfo{author}{\bibfnamefont{M.}~\bibnamefont{Kl\"{a}ui}},
  \bibinfo{author}{\bibfnamefont{Y.}~\bibnamefont{Mai}},
  \bibinfo{author}{\bibfnamefont{D.}~\bibnamefont{Beljonne}}, \bibnamefont{and}
  \bibinfo{author}{\bibfnamefont{H.~I.} \bibnamefont{Wang}},
  \bibinfo{journal}{Nano Lett.} \textbf{\bibinfo{volume}{20}},
  \bibinfo{pages}{2993} (\bibinfo{year}{2020}).

\bibitem[{\citenamefont{Schmidt et~al.}(2019)\citenamefont{Schmidt,
  Muruganathan, Kanzaki, Iwasaki, Hammam, Suzuki, Ogawa, and
  Mizuta}}]{schmidt2019dielectric}
\bibinfo{author}{\bibfnamefont{M.~E.} \bibnamefont{Schmidt}},
  \bibinfo{author}{\bibfnamefont{M.}~\bibnamefont{Muruganathan}},
  \bibinfo{author}{\bibfnamefont{T.}~\bibnamefont{Kanzaki}},
  \bibinfo{author}{\bibfnamefont{T.}~\bibnamefont{Iwasaki}},
  \bibinfo{author}{\bibfnamefont{A.~M.~M.} \bibnamefont{Hammam}},
  \bibinfo{author}{\bibfnamefont{S.}~\bibnamefont{Suzuki}},
  \bibinfo{author}{\bibfnamefont{S.}~\bibnamefont{Ogawa}}, \bibnamefont{and}
  \bibinfo{author}{\bibfnamefont{H.}~\bibnamefont{Mizuta}},
  \bibinfo{journal}{Small} \textbf{\bibinfo{volume}{15}},
  \bibinfo{pages}{1903025} (\bibinfo{year}{2019}).

\bibitem[{\citenamefont{Chazalviel}(1999)}]{chazalviel1999}
\bibinfo{author}{\bibfnamefont{J.-N.} \bibnamefont{Chazalviel}},
  \emph{\bibinfo{title}{Coulomb Screening by Mobile Charges: Applications to
  Materials Science, Chemistry, and Biology}} (\bibinfo{publisher}{Springer
  Science \& Business Media}, \bibinfo{year}{1999}).

\bibitem[{\citenamefont{Wang et~al.}(2020)\citenamefont{Wang, Liu, He, Zhang,
  Zhang, and Xie}}]{wang2020}
\bibinfo{author}{\bibfnamefont{H.}~\bibnamefont{Wang}},
  \bibinfo{author}{\bibfnamefont{W.}~\bibnamefont{Liu}},
  \bibinfo{author}{\bibfnamefont{X.}~\bibnamefont{He}},
  \bibinfo{author}{\bibfnamefont{P.}~\bibnamefont{Zhang}},
  \bibinfo{author}{\bibfnamefont{X.}~\bibnamefont{Zhang}}, \bibnamefont{and}
  \bibinfo{author}{\bibfnamefont{Y.}~\bibnamefont{Xie}}, \bibinfo{journal}{J.
  Am. Chem. Soc.} \textbf{\bibinfo{volume}{142}}, \bibinfo{pages}{14007}
  (\bibinfo{year}{2020}).

\bibitem[{\citenamefont{Dvorak et~al.}(2013)\citenamefont{Dvorak, Wei, and
  Wu}}]{dvorak2013}
\bibinfo{author}{\bibfnamefont{M.}~\bibnamefont{Dvorak}},
  \bibinfo{author}{\bibfnamefont{S.-H.} \bibnamefont{Wei}}, \bibnamefont{and}
  \bibinfo{author}{\bibfnamefont{Z.}~\bibnamefont{Wu}}, \bibinfo{journal}{Phys.
  Rev. Lett.} \textbf{\bibinfo{volume}{110}}, \bibinfo{pages}{016402}
  (\bibinfo{year}{2013}).

\bibitem[{\citenamefont{Cardona and Peter}(2005)}]{cardona2005}
\bibinfo{author}{\bibfnamefont{M.}~\bibnamefont{Cardona}} \bibnamefont{and}
  \bibinfo{author}{\bibfnamefont{Y.~Y.} \bibnamefont{Peter}},
  \emph{\bibinfo{title}{Fundamentals of semiconductors}}, vol.
  \bibinfo{volume}{619} (\bibinfo{publisher}{Springer}, \bibinfo{year}{2005}).

\bibitem[{\citenamefont{Chernikov et~al.}(2014)\citenamefont{Chernikov,
  Berkelbach, Hill, Rigosi, Li, Aslan, Reichman, Hybertsen, and
  Heinz}}]{chernikov2014}
\bibinfo{author}{\bibfnamefont{A.}~\bibnamefont{Chernikov}},
  \bibinfo{author}{\bibfnamefont{T.~C.} \bibnamefont{Berkelbach}},
  \bibinfo{author}{\bibfnamefont{H.~M.} \bibnamefont{Hill}},
  \bibinfo{author}{\bibfnamefont{A.}~\bibnamefont{Rigosi}},
  \bibinfo{author}{\bibfnamefont{Y.}~\bibnamefont{Li}},
  \bibinfo{author}{\bibfnamefont{B.}~\bibnamefont{Aslan}},
  \bibinfo{author}{\bibfnamefont{D.~R.} \bibnamefont{Reichman}},
  \bibinfo{author}{\bibfnamefont{M.~S.} \bibnamefont{Hybertsen}},
  \bibnamefont{and} \bibinfo{author}{\bibfnamefont{T.~F.} \bibnamefont{Heinz}},
  \bibinfo{journal}{Phys. Rev. Lett.} \textbf{\bibinfo{volume}{113}},
  \bibinfo{pages}{076802} (\bibinfo{year}{2014}).

\bibitem[{\citenamefont{Rohlfing and Louie}(2000)}]{rohlfing2000}
\bibinfo{author}{\bibfnamefont{M.}~\bibnamefont{Rohlfing}} \bibnamefont{and}
  \bibinfo{author}{\bibfnamefont{S.~G.} \bibnamefont{Louie}},
  \bibinfo{journal}{Phys. Rev. B} \textbf{\bibinfo{volume}{62}},
  \bibinfo{pages}{4927} (\bibinfo{year}{2000}).

\bibitem[{\citenamefont{Blase et~al.}(2018)\citenamefont{Blase, Duchemin, and
  Jacquemin}}]{blase2018}
\bibinfo{author}{\bibfnamefont{X.}~\bibnamefont{Blase}},
  \bibinfo{author}{\bibfnamefont{I.}~\bibnamefont{Duchemin}}, \bibnamefont{and}
  \bibinfo{author}{\bibfnamefont{D.}~\bibnamefont{Jacquemin}},
  \bibinfo{journal}{Chem. Soc. Rev.} \textbf{\bibinfo{volume}{47}},
  \bibinfo{pages}{1022} (\bibinfo{year}{2018}).

\bibitem[{\citenamefont{Molas et~al.}(2019)\citenamefont{Molas, Slobodeniuk,
  Nogajewski, Bartos, Babi{\'n}ski, Watanabe, Taniguchi, Faugeras, and
  Potemski}}]{molas2019energy}
\bibinfo{author}{\bibfnamefont{M.~R.} \bibnamefont{Molas}},
  \bibinfo{author}{\bibfnamefont{A.~O.} \bibnamefont{Slobodeniuk}},
  \bibinfo{author}{\bibfnamefont{K.}~\bibnamefont{Nogajewski}},
  \bibinfo{author}{\bibfnamefont{M.}~\bibnamefont{Bartos}},
  \bibinfo{author}{\bibfnamefont{A.}~\bibnamefont{Babi{\'n}ski}},
  \bibinfo{author}{\bibfnamefont{K.}~\bibnamefont{Watanabe}},
  \bibinfo{author}{\bibfnamefont{T.}~\bibnamefont{Taniguchi}},
  \bibinfo{author}{\bibfnamefont{C.}~\bibnamefont{Faugeras}}, \bibnamefont{and}
  \bibinfo{author}{\bibfnamefont{M.}~\bibnamefont{Potemski}},
  \bibinfo{journal}{Phys. Rev. Lett.} \textbf{\bibinfo{volume}{123}},
  \bibinfo{pages}{136801} (\bibinfo{year}{2019}).

\bibitem[{\citenamefont{Huang et~al.}(2013)\citenamefont{Huang, Liang, and
  Yang}}]{huang2013}
\bibinfo{author}{\bibfnamefont{S.}~\bibnamefont{Huang}},
  \bibinfo{author}{\bibfnamefont{Y.}~\bibnamefont{Liang}}, \bibnamefont{and}
  \bibinfo{author}{\bibfnamefont{L.}~\bibnamefont{Yang}},
  \bibinfo{journal}{Phys. Rev. B} \textbf{\bibinfo{volume}{88}},
  \bibinfo{pages}{075441} (\bibinfo{year}{2013}).

\bibitem[{\citenamefont{Villegas et~al.}(2016)\citenamefont{Villegas, Rodin,
  Carvalho, and Rocha}}]{villegas2016}
\bibinfo{author}{\bibfnamefont{C.~E.~P.} \bibnamefont{Villegas}},
  \bibinfo{author}{\bibfnamefont{A.~S.} \bibnamefont{Rodin}},
  \bibinfo{author}{\bibfnamefont{A.}~\bibnamefont{Carvalho}}, \bibnamefont{and}
  \bibinfo{author}{\bibfnamefont{A.~R.} \bibnamefont{Rocha}},
  \bibinfo{journal}{Phys. Chem. Chem. Phys.} \textbf{\bibinfo{volume}{18}},
  \bibinfo{pages}{27829} (\bibinfo{year}{2016}).

\bibitem[{\citenamefont{Martins~Quintela and Peres}(2020)}]{martins2020}
\bibinfo{author}{\bibfnamefont{M.~F.~C.} \bibnamefont{Martins~Quintela}}
  \bibnamefont{and} \bibinfo{author}{\bibfnamefont{N.~M.~R.}
  \bibnamefont{Peres}}, \bibinfo{journal}{Eur. Phys. J. B}
  \textbf{\bibinfo{volume}{93}}, \bibinfo{pages}{1} (\bibinfo{year}{2020}).

\bibitem[{\citenamefont{Rytova}(1967)}]{rytova1967screened}
\bibinfo{author}{\bibfnamefont{N.~S.} \bibnamefont{Rytova}},
  \bibinfo{journal}{Mosc. Univ. Phys. Bull.} \textbf{\bibinfo{volume}{3}},
  \bibinfo{pages}{30} (\bibinfo{year}{1967}).

\bibitem[{\citenamefont{Keldysh}(1979)}]{keldysh1979}
\bibinfo{author}{\bibfnamefont{L.~V.} \bibnamefont{Keldysh}},
  \bibinfo{journal}{Sov. J. Exp. Theor. Phys. Lett.}
  \textbf{\bibinfo{volume}{29}}, \bibinfo{pages}{658} (\bibinfo{year}{1979}).

\bibitem[{\citenamefont{Cudazzo et~al.}(2011)\citenamefont{Cudazzo, Tokatly,
  and Rubio}}]{cudazzo2011}
\bibinfo{author}{\bibfnamefont{P.}~\bibnamefont{Cudazzo}},
  \bibinfo{author}{\bibfnamefont{I.~V.} \bibnamefont{Tokatly}},
  \bibnamefont{and} \bibinfo{author}{\bibfnamefont{A.}~\bibnamefont{Rubio}},
  \bibinfo{journal}{Phys. Rev. B} \textbf{\bibinfo{volume}{84}},
  \bibinfo{pages}{085406} (\bibinfo{year}{2011}).

\bibitem[{\citenamefont{Ogawa and Takagahara}(1991)}]{ogawa1991optical}
\bibinfo{author}{\bibfnamefont{T.}~\bibnamefont{Ogawa}} \bibnamefont{and}
  \bibinfo{author}{\bibfnamefont{T.}~\bibnamefont{Takagahara}},
  \bibinfo{journal}{Phys. Rev. B} \textbf{\bibinfo{volume}{44}},
  \bibinfo{pages}{8138} (\bibinfo{year}{1991}).

\bibitem[{\citenamefont{Wang et~al.}(2014)\citenamefont{Wang, Miao, and
  Zhai}}]{wang2014}
\bibinfo{author}{\bibfnamefont{Y.}~\bibnamefont{Wang}},
  \bibinfo{author}{\bibfnamefont{W.-D.} \bibnamefont{Miao}}, \bibnamefont{and}
  \bibinfo{author}{\bibfnamefont{L.-X.} \bibnamefont{Zhai}},
  \bibinfo{journal}{Phys. Lett. A} \textbf{\bibinfo{volume}{378}},
  \bibinfo{pages}{442} (\bibinfo{year}{2014}).

\bibitem[{\citenamefont{Bryant and Glick}(1982)}]{bryant1982}
\bibinfo{author}{\bibfnamefont{G.~W.} \bibnamefont{Bryant}} \bibnamefont{and}
  \bibinfo{author}{\bibfnamefont{A.~J.} \bibnamefont{Glick}},
  \bibinfo{journal}{Phys. Rev. B} \textbf{\bibinfo{volume}{26}},
  \bibinfo{pages}{5855} (\bibinfo{year}{1982}).

\bibitem[{\citenamefont{Brazovskii and Kirova}(2010)}]{brazovskii2010physical}
\bibinfo{author}{\bibfnamefont{S.}~\bibnamefont{Brazovskii}} \bibnamefont{and}
  \bibinfo{author}{\bibfnamefont{N.}~\bibnamefont{Kirova}},
  \bibinfo{journal}{Chem. Soc. Rev.} \textbf{\bibinfo{volume}{39}},
  \bibinfo{pages}{2453} (\bibinfo{year}{2010}).

\bibitem[{\citenamefont{Grasselli}(2017)}]{grasselli2017}
\bibinfo{author}{\bibfnamefont{F.}~\bibnamefont{Grasselli}},
  \bibinfo{journal}{Am. J. Phys.} \textbf{\bibinfo{volume}{85}},
  \bibinfo{pages}{834} (\bibinfo{year}{2017}).

\bibitem[{\citenamefont{Deslippe et~al.}(2009)\citenamefont{Deslippe, Dipoppa,
  Prendergast, Moutinho, Capaz, and Louie}}]{deslippe2009}
\bibinfo{author}{\bibfnamefont{J.}~\bibnamefont{Deslippe}},
  \bibinfo{author}{\bibfnamefont{M.}~\bibnamefont{Dipoppa}},
  \bibinfo{author}{\bibfnamefont{D.}~\bibnamefont{Prendergast}},
  \bibinfo{author}{\bibfnamefont{M.~V.~O.} \bibnamefont{Moutinho}},
  \bibinfo{author}{\bibfnamefont{R.~B.} \bibnamefont{Capaz}}, \bibnamefont{and}
  \bibinfo{author}{\bibfnamefont{S.~G.} \bibnamefont{Louie}},
  \bibinfo{journal}{Nano Lett.} \textbf{\bibinfo{volume}{9}},
  \bibinfo{pages}{1330} (\bibinfo{year}{2009}).

\bibitem[{\citenamefont{Sesti et~al.}(2022)\citenamefont{Sesti, Varsano,
  Molinari, and Rontani}}]{sesti2022}
\bibinfo{author}{\bibfnamefont{G.}~\bibnamefont{Sesti}},
  \bibinfo{author}{\bibfnamefont{D.}~\bibnamefont{Varsano}},
  \bibinfo{author}{\bibfnamefont{E.}~\bibnamefont{Molinari}}, \bibnamefont{and}
  \bibinfo{author}{\bibfnamefont{M.}~\bibnamefont{Rontani}},
  \bibinfo{journal}{Phys. Rev. B} \textbf{\bibinfo{volume}{105}},
  \bibinfo{pages}{195404} (\bibinfo{year}{2022}).

\bibitem[{\citenamefont{Cai et~al.}(2010)\citenamefont{Cai, Ruffieux, Jaafar,
  Bieri, Braun, Blankenburg, Muoth, Seitsonen, Saleh, Feng et~al.}}]{cai2010}
\bibinfo{author}{\bibfnamefont{J.}~\bibnamefont{Cai}},
  \bibinfo{author}{\bibfnamefont{P.}~\bibnamefont{Ruffieux}},
  \bibinfo{author}{\bibfnamefont{R.}~\bibnamefont{Jaafar}},
  \bibinfo{author}{\bibfnamefont{M.}~\bibnamefont{Bieri}},
  \bibinfo{author}{\bibfnamefont{T.}~\bibnamefont{Braun}},
  \bibinfo{author}{\bibfnamefont{S.}~\bibnamefont{Blankenburg}},
  \bibinfo{author}{\bibfnamefont{M.}~\bibnamefont{Muoth}},
  \bibinfo{author}{\bibfnamefont{A.~P.} \bibnamefont{Seitsonen}},
  \bibinfo{author}{\bibfnamefont{M.}~\bibnamefont{Saleh}},
  \bibinfo{author}{\bibfnamefont{X.}~\bibnamefont{Feng}}, \bibnamefont{et~al.},
  \bibinfo{journal}{Nature} \textbf{\bibinfo{volume}{466}},
  \bibinfo{pages}{470} (\bibinfo{year}{2010}).

\bibitem[{\citenamefont{Bulashevich et~al.}(2003)\citenamefont{Bulashevich,
  Suris, and Rotkin}}]{bulashevich2003}
\bibinfo{author}{\bibfnamefont{K.~A.} \bibnamefont{Bulashevich}},
  \bibinfo{author}{\bibfnamefont{R.~A.} \bibnamefont{Suris}}, \bibnamefont{and}
  \bibinfo{author}{\bibfnamefont{S.~V.} \bibnamefont{Rotkin}},
  \bibinfo{journal}{Int. J. Nanosci.} \textbf{\bibinfo{volume}{2}},
  \bibinfo{pages}{521} (\bibinfo{year}{2003}).

\bibitem[{\citenamefont{Wang et~al.}(2011)\citenamefont{Wang, Chen, and
  Wang}}]{wang2011optical}
\bibinfo{author}{\bibfnamefont{S.}~\bibnamefont{Wang}},
  \bibinfo{author}{\bibfnamefont{Q.}~\bibnamefont{Chen}}, \bibnamefont{and}
  \bibinfo{author}{\bibfnamefont{J.}~\bibnamefont{Wang}},
  \bibinfo{journal}{App. Phys. Lett.} \textbf{\bibinfo{volume}{99}}
  (\bibinfo{year}{2011}).

\bibitem[{\citenamefont{Wang et~al.}(2016)\citenamefont{Wang, Zhou, Li, Li, Wu,
  Duan, and He}}]{wang2016energy}
\bibinfo{author}{\bibfnamefont{W.-X.} \bibnamefont{Wang}},
  \bibinfo{author}{\bibfnamefont{M.}~\bibnamefont{Zhou}},
  \bibinfo{author}{\bibfnamefont{X.}~\bibnamefont{Li}},
  \bibinfo{author}{\bibfnamefont{S.-Y.} \bibnamefont{Li}},
  \bibinfo{author}{\bibfnamefont{X.}~\bibnamefont{Wu}},
  \bibinfo{author}{\bibfnamefont{W.}~\bibnamefont{Duan}}, \bibnamefont{and}
  \bibinfo{author}{\bibfnamefont{L.}~\bibnamefont{He}}, \bibinfo{journal}{Phys.
  Rev, B} \textbf{\bibinfo{volume}{93}}, \bibinfo{pages}{241403}
  (\bibinfo{year}{2016}).

\bibitem[{\citenamefont{Olsen et~al.}(2016)\citenamefont{Olsen, Latini,
  Rasmussen, and Thygesen}}]{olsen2016}
\bibinfo{author}{\bibfnamefont{T.}~\bibnamefont{Olsen}},
  \bibinfo{author}{\bibfnamefont{S.}~\bibnamefont{Latini}},
  \bibinfo{author}{\bibfnamefont{F.}~\bibnamefont{Rasmussen}},
  \bibnamefont{and} \bibinfo{author}{\bibfnamefont{K.~S.}
  \bibnamefont{Thygesen}}, \bibinfo{journal}{Phys. Rev. Lett.}
  \textbf{\bibinfo{volume}{116}}, \bibinfo{pages}{056401}
  (\bibinfo{year}{2016}).

\bibitem[{\citenamefont{Jiang et~al.}(2017)\citenamefont{Jiang, Liu, Li, and
  Duan}}]{jiang2017}
\bibinfo{author}{\bibfnamefont{Z.}~\bibnamefont{Jiang}},
  \bibinfo{author}{\bibfnamefont{Z.}~\bibnamefont{Liu}},
  \bibinfo{author}{\bibfnamefont{Y.}~\bibnamefont{Li}}, \bibnamefont{and}
  \bibinfo{author}{\bibfnamefont{W.}~\bibnamefont{Duan}},
  \bibinfo{journal}{Phys. Rev. Lett.} \textbf{\bibinfo{volume}{118}},
  \bibinfo{pages}{266401} (\bibinfo{year}{2017}).

\bibitem[{\citenamefont{Van Den~Brink and Sawatzky}(2000)}]{van2000}
\bibinfo{author}{\bibfnamefont{J.}~\bibnamefont{Van Den~Brink}}
  \bibnamefont{and} \bibinfo{author}{\bibfnamefont{G.~A.}
  \bibnamefont{Sawatzky}}, \bibinfo{journal}{Europhys. Lett.}
  \textbf{\bibinfo{volume}{50}}, \bibinfo{pages}{447} (\bibinfo{year}{2000}).

\bibitem[{sup()}]{supp}
\bibinfo{note}{The supplementary material online contains the description of
  the methodology, parameters used, convergence tests, and the dependence of
  the effective Coulomb force for AGNRs of different widths. The Supplemental
  material also contains Refs.
  \cite{KohnSham1965,HohenbergKohn1964,qe,van2018,pbe,yambo,bruneval2008,wang2011optical,marinho2024photovoltaic,perez2020}}.

\bibitem[{\citenamefont{Williams et~al.}(2011)\citenamefont{Williams, Low,
  Lundstrom, and Marcus}}]{williams2011gate}
\bibinfo{author}{\bibfnamefont{J.~R.} \bibnamefont{Williams}},
  \bibinfo{author}{\bibfnamefont{T.}~\bibnamefont{Low}},
  \bibinfo{author}{\bibfnamefont{M.~S.} \bibnamefont{Lundstrom}},
  \bibnamefont{and} \bibinfo{author}{\bibfnamefont{C.~M.}
  \bibnamefont{Marcus}}, \bibinfo{journal}{Nat. Nanotechnol.}
  \textbf{\bibinfo{volume}{6}}, \bibinfo{pages}{222} (\bibinfo{year}{2011}).

\bibitem[{\citenamefont{Zhang et~al.}(2009)\citenamefont{Zhang, He, and
  Chen}}]{zhang2009guided}
\bibinfo{author}{\bibfnamefont{F.-M.} \bibnamefont{Zhang}},
  \bibinfo{author}{\bibfnamefont{Y.}~\bibnamefont{He}}, \bibnamefont{and}
  \bibinfo{author}{\bibfnamefont{X.}~\bibnamefont{Chen}},
  \bibinfo{journal}{Appl. Phys. Lett.} \textbf{\bibinfo{volume}{94}}
  (\bibinfo{year}{2009}).

\bibitem[{\citenamefont{Villegas and Tavares}(2010)}]{villegas2010comment}
\bibinfo{author}{\bibfnamefont{C.~E.~P.} \bibnamefont{Villegas}}
  \bibnamefont{and} \bibinfo{author}{\bibfnamefont{M.~R.~S.}
  \bibnamefont{Tavares}}, \bibinfo{journal}{Appl. Phys. Lett.}
  \textbf{\bibinfo{volume}{96}}, \bibinfo{pages}{186101}
  (\bibinfo{year}{2010}).

\bibitem[{\citenamefont{Villegas and Tavares}(2012)}]{villegas2012}
\bibinfo{author}{\bibfnamefont{C.~E.~P.} \bibnamefont{Villegas}}
  \bibnamefont{and} \bibinfo{author}{\bibfnamefont{M.~R.~S.}
  \bibnamefont{Tavares}}, \bibinfo{journal}{Appl. Phys. Lett.}
  \textbf{\bibinfo{volume}{101}} (\bibinfo{year}{2012}).

\bibitem[{\citenamefont{Villegas et~al.}(2013)\citenamefont{Villegas, Tavares,
  Hai, and Vasilopoulos}}]{villegas2013}
\bibinfo{author}{\bibfnamefont{C.~E.~P.} \bibnamefont{Villegas}},
  \bibinfo{author}{\bibfnamefont{M.~R.~S.} \bibnamefont{Tavares}},
  \bibinfo{author}{\bibfnamefont{G.-Q.} \bibnamefont{Hai}}, \bibnamefont{and}
  \bibinfo{author}{\bibfnamefont{P.}~\bibnamefont{Vasilopoulos}},
  \bibinfo{journal}{Phys. Rev. B} \textbf{\bibinfo{volume}{88}},
  \bibinfo{pages}{165426} (\bibinfo{year}{2013}).

\bibitem[{\citenamefont{Lighthill et~al.}(1958)\citenamefont{Lighthill,
  Lighthill, Lighthill et~al.}}]{lighthill1958}
\bibinfo{author}{\bibfnamefont{M.~J.} \bibnamefont{Lighthill}},
  \bibinfo{author}{\bibfnamefont{M.~J.~S.} \bibnamefont{Lighthill}},
  \bibinfo{author}{\bibfnamefont{M.~J.} \bibnamefont{Lighthill}},
  \bibnamefont{et~al.}, \emph{\bibinfo{title}{An introduction to Fourier
  analysis and generalised functions}} (\bibinfo{publisher}{Cambridge
  University Press}, \bibinfo{year}{1958}).

\bibitem[{\citenamefont{Sarma and Hwang}(2009)}]{dassarma2009}
\bibinfo{author}{\bibfnamefont{S.~D.} \bibnamefont{Sarma}} \bibnamefont{and}
  \bibinfo{author}{\bibfnamefont{E.~H.} \bibnamefont{Hwang}},
  \bibinfo{journal}{Phys. Rev. Lett.} \textbf{\bibinfo{volume}{102}},
  \bibinfo{pages}{206412} (\bibinfo{year}{2009}).

\bibitem[{\citenamefont{Thakur et~al.}(2017)\citenamefont{Thakur, Sachdeva, and
  Agarwal}}]{thakur2017dynamical}
\bibinfo{author}{\bibfnamefont{A.}~\bibnamefont{Thakur}},
  \bibinfo{author}{\bibfnamefont{R.}~\bibnamefont{Sachdeva}}, \bibnamefont{and}
  \bibinfo{author}{\bibfnamefont{A.}~\bibnamefont{Agarwal}},
  \bibinfo{journal}{J. Phys. Condens. Matter.} \textbf{\bibinfo{volume}{29}},
  \bibinfo{pages}{105701} (\bibinfo{year}{2017}).

\bibitem[{\citenamefont{Haug and Koch}(2009)}]{haug2009quantum}
\bibinfo{author}{\bibfnamefont{H.}~\bibnamefont{Haug}} \bibnamefont{and}
  \bibinfo{author}{\bibfnamefont{S.~W.} \bibnamefont{Koch}},
  \emph{\bibinfo{title}{Quantum theory of the optical and electronic properties
  of semiconductors}} (\bibinfo{publisher}{World Scientific Publishing
  Company}, \bibinfo{year}{2009}).

\bibitem[{\citenamefont{Wannier}(1937)}]{wannier1937}
\bibinfo{author}{\bibfnamefont{G.~H.} \bibnamefont{Wannier}},
  \bibinfo{journal}{Phys. Rev.} \textbf{\bibinfo{volume}{52}},
  \bibinfo{pages}{191} (\bibinfo{year}{1937}).

\bibitem[{\citenamefont{Capaz et~al.}(2006)\citenamefont{Capaz, Spataru,
  Ismail-Beigi, and Louie}}]{capaz2006diameter}
\bibinfo{author}{\bibfnamefont{R.~B.} \bibnamefont{Capaz}},
  \bibinfo{author}{\bibfnamefont{C.~D.} \bibnamefont{Spataru}},
  \bibinfo{author}{\bibfnamefont{S.}~\bibnamefont{Ismail-Beigi}},
  \bibnamefont{and} \bibinfo{author}{\bibfnamefont{S.~G.} \bibnamefont{Louie}},
  \bibinfo{journal}{Phys. Rev. B} \textbf{\bibinfo{volume}{74}},
  \bibinfo{pages}{121401} (\bibinfo{year}{2006}).

\bibitem[{\citenamefont{Shapir}(2019)}]{shapir2019imaging}
\bibinfo{author}{\bibfnamefont{I.}~\bibnamefont{Shapir}}, Ph.D. thesis,
  \bibinfo{school}{The Weizmann Institute of Science (Israel)}
  (\bibinfo{year}{2019}).

\bibitem[{\citenamefont{Zhu and Su}(2011)}]{zhu2011scaling}
\bibinfo{author}{\bibfnamefont{X.}~\bibnamefont{Zhu}} \bibnamefont{and}
  \bibinfo{author}{\bibfnamefont{H.}~\bibnamefont{Su}}, \bibinfo{journal}{J.
  Phys. Chem. A} \textbf{\bibinfo{volume}{115}}, \bibinfo{pages}{11998}
  (\bibinfo{year}{2011}).

\bibitem[{\citenamefont{Jorio et~al.}(2005)\citenamefont{Jorio, Fantini,
  Pimenta, Capaz, Samsonidze, Dresselhaus, Dresselhaus, Jiang, Kobayashi,
  Gr{\"u}neis et~al.}}]{jorio2005resonance}
\bibinfo{author}{\bibfnamefont{A.}~\bibnamefont{Jorio}},
  \bibinfo{author}{\bibfnamefont{C.}~\bibnamefont{Fantini}},
  \bibinfo{author}{\bibfnamefont{M.~A.} \bibnamefont{Pimenta}},
  \bibinfo{author}{\bibfnamefont{R.~B.} \bibnamefont{Capaz}},
  \bibinfo{author}{\bibfnamefont{G.~G.} \bibnamefont{Samsonidze}},
  \bibinfo{author}{\bibfnamefont{G.}~\bibnamefont{Dresselhaus}},
  \bibinfo{author}{\bibfnamefont{M.~S.} \bibnamefont{Dresselhaus}},
  \bibinfo{author}{\bibfnamefont{J.}~\bibnamefont{Jiang}},
  \bibinfo{author}{\bibfnamefont{N.}~\bibnamefont{Kobayashi}},
  \bibinfo{author}{\bibfnamefont{A.}~\bibnamefont{Gr{\"u}neis}},
  \bibnamefont{et~al.}, \bibinfo{journal}{Phys. Rev. B}
  \textbf{\bibinfo{volume}{71}}, \bibinfo{pages}{075401}
  (\bibinfo{year}{2005}).

\bibitem[{\citenamefont{Denk et~al.}(2014)\citenamefont{Denk, Hohage,
  Zeppenfeld, Cai, Pignedoli, S{\"o}de, Fasel, Feng, M{\"u}llen, Wang
  et~al.}}]{denk2014}
\bibinfo{author}{\bibfnamefont{R.}~\bibnamefont{Denk}},
  \bibinfo{author}{\bibfnamefont{M.}~\bibnamefont{Hohage}},
  \bibinfo{author}{\bibfnamefont{P.}~\bibnamefont{Zeppenfeld}},
  \bibinfo{author}{\bibfnamefont{J.}~\bibnamefont{Cai}},
  \bibinfo{author}{\bibfnamefont{C.~A.} \bibnamefont{Pignedoli}},
  \bibinfo{author}{\bibfnamefont{H.}~\bibnamefont{S{\"o}de}},
  \bibinfo{author}{\bibfnamefont{R.}~\bibnamefont{Fasel}},
  \bibinfo{author}{\bibfnamefont{X.}~\bibnamefont{Feng}},
  \bibinfo{author}{\bibfnamefont{K.}~\bibnamefont{M{\"u}llen}},
  \bibinfo{author}{\bibfnamefont{S.}~\bibnamefont{Wang}}, \bibnamefont{et~al.},
  \bibinfo{journal}{Nat. Commun.} \textbf{\bibinfo{volume}{5}},
  \bibinfo{pages}{1} (\bibinfo{year}{2014}).

\bibitem[{\citenamefont{Riis-Jensen et~al.}(2020)\citenamefont{Riis-Jensen,
  Gjerding, Russo, and Thygesen}}]{riis2020}
\bibinfo{author}{\bibfnamefont{A.~C.} \bibnamefont{Riis-Jensen}},
  \bibinfo{author}{\bibfnamefont{M.~N.} \bibnamefont{Gjerding}},
  \bibinfo{author}{\bibfnamefont{S.}~\bibnamefont{Russo}}, \bibnamefont{and}
  \bibinfo{author}{\bibfnamefont{K.~S.} \bibnamefont{Thygesen}},
  \bibinfo{journal}{Phys. Rev. B} \textbf{\bibinfo{volume}{102}},
  \bibinfo{pages}{201402} (\bibinfo{year}{2020}).

\bibitem[{\citenamefont{Mahan}(1998)}]{mahan}
\bibinfo{author}{\bibfnamefont{G.}~\bibnamefont{Mahan}},
  \emph{\bibinfo{title}{Many-Particle Physics}} (\bibinfo{publisher}{New York:
  Plenum}, \bibinfo{year}{1998}).

\bibitem[{\citenamefont{Li and Sarma}(1989)}]{li1989}
\bibinfo{author}{\bibfnamefont{Q.}~\bibnamefont{Li}} \bibnamefont{and}
  \bibinfo{author}{\bibfnamefont{S.~D.} \bibnamefont{Sarma}},
  \bibinfo{journal}{Phys. Rev. B} \textbf{\bibinfo{volume}{40}},
  \bibinfo{pages}{5860} (\bibinfo{year}{1989}).

\bibitem[{\citenamefont{Loudon}(2015)}]{loudon2015}
\bibinfo{author}{\bibfnamefont{R.}~\bibnamefont{Loudon}},
  \bibinfo{journal}{Proc. R. Soc. A} \textbf{\bibinfo{volume}{472}},
  \bibinfo{pages}{20150534} (\bibinfo{year}{2015}).

\bibitem[{\citenamefont{Rakhmanov et~al.}(2015)\citenamefont{Rakhmanov,
  Karpova, Rakhimboeva, Khashimova, and Babajanov}}]{rakhmanov2015}
\bibinfo{author}{\bibfnamefont{S.}~\bibnamefont{Rakhmanov}},
  \bibinfo{author}{\bibfnamefont{O.}~\bibnamefont{Karpova}},
  \bibinfo{author}{\bibfnamefont{D.~R.} \bibnamefont{Rakhimboeva}},
  \bibinfo{author}{\bibfnamefont{F.}~\bibnamefont{Khashimova}},
  \bibnamefont{and}
  \bibinfo{author}{\bibfnamefont{D.}~\bibnamefont{Babajanov}},
  \bibinfo{journal}{Nanosyst.: Phys. Chem. Math.} \textbf{\bibinfo{volume}{6}},
  \bibinfo{pages}{767} (\bibinfo{year}{2015}).

\bibitem[{\citenamefont{Kohn and Sham}(1965)}]{KohnSham1965}
\bibinfo{author}{\bibfnamefont{W.}~\bibnamefont{Kohn}} \bibnamefont{and}
  \bibinfo{author}{\bibfnamefont{L.~J.} \bibnamefont{Sham}},
  \bibinfo{journal}{Phys. Rev.} \textbf{\bibinfo{volume}{140}},
  \bibinfo{pages}{A1133} (\bibinfo{year}{1965}).

\bibitem[{\citenamefont{Hohenberg and Kohn}(1964)}]{HohenbergKohn1964}
\bibinfo{author}{\bibfnamefont{P.}~\bibnamefont{Hohenberg}} \bibnamefont{and}
  \bibinfo{author}{\bibfnamefont{W.}~\bibnamefont{Kohn}},
  \bibinfo{journal}{Phys. Rev.} \textbf{\bibinfo{volume}{136}},
  \bibinfo{pages}{B864} (\bibinfo{year}{1964}).

\bibitem[{\citenamefont{Giannozzi et~al.}(2009)\citenamefont{Giannozzi, Baroni,
  Bonini, Calandra, Car, Cavazzoni, Ceresoli, Chiarotti, Cococcioni, Dabo
  et~al.}}]{qe}
\bibinfo{author}{\bibfnamefont{P.}~\bibnamefont{Giannozzi}},
  \bibinfo{author}{\bibfnamefont{S.}~\bibnamefont{Baroni}},
  \bibinfo{author}{\bibfnamefont{N.}~\bibnamefont{Bonini}},
  \bibinfo{author}{\bibfnamefont{M.}~\bibnamefont{Calandra}},
  \bibinfo{author}{\bibfnamefont{R.}~\bibnamefont{Car}},
  \bibinfo{author}{\bibfnamefont{C.}~\bibnamefont{Cavazzoni}},
  \bibinfo{author}{\bibfnamefont{D.}~\bibnamefont{Ceresoli}},
  \bibinfo{author}{\bibfnamefont{G.~L.} \bibnamefont{Chiarotti}},
  \bibinfo{author}{\bibfnamefont{M.}~\bibnamefont{Cococcioni}},
  \bibinfo{author}{\bibfnamefont{I.}~\bibnamefont{Dabo}}, \bibnamefont{et~al.},
  \bibinfo{journal}{J. Phys.: Condens. Matter} \textbf{\bibinfo{volume}{21}},
  \bibinfo{pages}{395502} (\bibinfo{year}{2009}).

\bibitem[{\citenamefont{van Setten et~al.}(2018)\citenamefont{van Setten,
  Giantomassi, Bousquet, Verstraete, Hamann, Gonze, and Rignanese}}]{van2018}
\bibinfo{author}{\bibfnamefont{M.~J.} \bibnamefont{van Setten}},
  \bibinfo{author}{\bibfnamefont{M.}~\bibnamefont{Giantomassi}},
  \bibinfo{author}{\bibfnamefont{E.}~\bibnamefont{Bousquet}},
  \bibinfo{author}{\bibfnamefont{M.~J.} \bibnamefont{Verstraete}},
  \bibinfo{author}{\bibfnamefont{D.~R.} \bibnamefont{Hamann}},
  \bibinfo{author}{\bibfnamefont{X.}~\bibnamefont{Gonze}}, \bibnamefont{and}
  \bibinfo{author}{\bibfnamefont{G.-M.} \bibnamefont{Rignanese}},
  \bibinfo{journal}{Comput. Phys. Commun.} \textbf{\bibinfo{volume}{226}},
  \bibinfo{pages}{39} (\bibinfo{year}{2018}).

\bibitem[{\citenamefont{Perdew et~al.}(1996)\citenamefont{Perdew, Burke, and
  Ernzerhof}}]{pbe}
\bibinfo{author}{\bibfnamefont{J.~P.} \bibnamefont{Perdew}},
  \bibinfo{author}{\bibfnamefont{K.}~\bibnamefont{Burke}}, \bibnamefont{and}
  \bibinfo{author}{\bibfnamefont{M.}~\bibnamefont{Ernzerhof}},
  \bibinfo{journal}{Phys. Rev. Lett.} \textbf{\bibinfo{volume}{77}},
  \bibinfo{pages}{3865} (\bibinfo{year}{1996}).

\bibitem[{\citenamefont{Marini et~al.}(2009)\citenamefont{Marini, Hogan,
  Gr\"{u}ning, and Varsano}}]{yambo}
\bibinfo{author}{\bibfnamefont{A.}~\bibnamefont{Marini}},
  \bibinfo{author}{\bibfnamefont{C.}~\bibnamefont{Hogan}},
  \bibinfo{author}{\bibfnamefont{M.}~\bibnamefont{Gr\"{u}ning}},
  \bibnamefont{and} \bibinfo{author}{\bibfnamefont{D.}~\bibnamefont{Varsano}},
  \bibinfo{journal}{Comput. Phys. Commun.} \textbf{\bibinfo{volume}{180}},
  \bibinfo{pages}{1392} (\bibinfo{year}{2009}), ISSN \bibinfo{issn}{0010-4655}.

\bibitem[{\citenamefont{Bruneval and Gonze}(2008)}]{bruneval2008}
\bibinfo{author}{\bibfnamefont{F.}~\bibnamefont{Bruneval}} \bibnamefont{and}
  \bibinfo{author}{\bibfnamefont{X.}~\bibnamefont{Gonze}},
  \bibinfo{journal}{Phys. Rev. B} \textbf{\bibinfo{volume}{78}},
  \bibinfo{pages}{085125} (\bibinfo{year}{2008}).

\bibitem[{\citenamefont{Marinho~Jr et~al.}(2024)\citenamefont{Marinho~Jr,
  Villegas, Venezuela, and Rocha}}]{marinho2024photovoltaic}
\bibinfo{author}{\bibfnamefont{E.}~\bibnamefont{Marinho~Jr}},
  \bibinfo{author}{\bibfnamefont{C.~E.~P.} \bibnamefont{Villegas}},
  \bibinfo{author}{\bibfnamefont{P.}~\bibnamefont{Venezuela}},
  \bibnamefont{and} \bibinfo{author}{\bibfnamefont{A.~R.} \bibnamefont{Rocha}},
  \bibinfo{journal}{ACS Appl. Energy Mater.} \textbf{\bibinfo{volume}{7}},
  \bibinfo{pages}{1051–1059} (\bibinfo{year}{2024}).

\bibitem[{\citenamefont{Perez et~al.}(2020)\citenamefont{Perez, Amorim,
  Villegas, and Rocha}}]{perez2020}
\bibinfo{author}{\bibfnamefont{A.}~\bibnamefont{Perez}},
  \bibinfo{author}{\bibfnamefont{R.~G.} \bibnamefont{Amorim}},
  \bibinfo{author}{\bibfnamefont{C.~E.~P.} \bibnamefont{Villegas}},
  \bibnamefont{and} \bibinfo{author}{\bibfnamefont{A.~R.} \bibnamefont{Rocha}},
  \bibinfo{journal}{Phys. Chem. Chem. Phys.} \textbf{\bibinfo{volume}{22}},
  \bibinfo{pages}{27053} (\bibinfo{year}{2020}).

\end{thebibliography}

\end{document}